\documentclass[preprint2]{aastex} 
\usepackage{graphicx}
\usepackage{amsmath}
\usepackage{natbib}
\usepackage[colorlinks=true,citecolor=blue,breaklinks=true,linktocpage=true]{hyperref}
\bibpunct{(}{)}{;}{a}{}{,} 
\usepackage[switch,pagewise]{lineno}
\usepackage{xspace}
\usepackage{xparse} 
\usepackage{xcolor} 

\newcommand{\alfven}{Alfv\'en\xspace}
\newcommand{\unit}[1]{\ensuremath{\,\mathrm {#1}}}  
\renewcommand{\d}{\ensuremath{\text{d}}}  
\newcommand{\A}{\ensuremath{\text{A}}}  %
\newcommand{\s}{\ensuremath{\text{s}}} 
\newcommand{\T}{\ensuremath{\text{T}}}  
\newcommand{\D}{\ensuremath{\text{D}}} 
\renewcommand{\i}{\ensuremath{\text{i}}}  
\newcommand{\e}{\ensuremath{\text{e}}} 
\DeclareDocumentCommand\figref{ m g }{{Figure~\ref{#1}\IfNoValueF {#2} {(#2)}}}
\newcommand{\secref}[1]{Section~\ref{#1}}
\newcommand{\tabref}[1]{Table~\ref{#1}}
\renewcommand{\eqref}[1]{Equation~\ref{#1}}
\newcommand{\bvec}[1]{\ensuremath{\boldsymbol{\mathbf{#1}}}}  
\renewcommand{\div}[1]{\nabla\cdot #1} 
\newcommand{\curl}[1]{\nabla\times #1} 
\newcommand{\grad}[1]{\nabla#1} 
\newcommand{\RR}{\ensuremath{\mathcal{R}}}
\newcommand{\singlequote}[1]{\lq{#1}\rq}
\newcommand{\feline}{\ensuremath{\lambda1118.1\unit{\AA{}}}\xspace}
\renewcommand{\deg}{\ensuremath{^\circ}}

\shorttitle{standing slow wave}
\shortauthors{Yuan et al.}

\begin{document}

\title{Forward Modelling of Standing Slow Modes in Flaring Coronal Loops}
\author{D. Yuan\altaffilmark{1,2}}
\email{Ding.Yuan@wis.kuleuven.be}
\author{T. Van Doorsselaere\altaffilmark{1}}
\author{D. Banerjee\altaffilmark{3}}
\and
\author{P. Antolin\altaffilmark{4}}

\altaffiltext{1}{Centre for mathematical Plasma Astrophysics, Department of Mathematics, KU Leuven, Celestijnenlaan 200B bus 2400, B-3001 Leuven, Belgium}
\altaffiltext{2}{Key Laboratory of Solar Activity, National Astronomical Observatories, Chinese Academy of Sciences, Beijing, 100012}
\altaffiltext{3}{Indian Institute of Astrophysics, II Block, Koramangala, Bangalore 560 034, India}
\altaffiltext{4}{National Astronomical Observatory of Japan, 2-21-1 Osawa, Mitaka, Tokyo 181-8588, Japan}
\begin{abstract}
Standing slow mode waves in hot flaring loops are exclusively observed in spectrometers and are used to diagnose the magnetic field strength and temperature of the loop structure. Due to the lack of spatial information, the longitudinal mode cannot be effectively identified. In this study, we simulate standing slow mode waves in flaring loops and compare the synthesized line emission properties with SUMER spectrographic and SDO/AIA imaging observations. We find that the emission intensity and line width oscillations are a quarter period out of phase with Doppler shift velocity both in time and spatial domain, which can be used to identify a standing slow mode wave from spectroscopic observations. However, the longitudinal overtones could be only measured with the assistance of imagers. We find emission intensity asymmetry in the positive and negative modulations, this is because the contribution function pertaining to the atomic emission process responds differently to positive and negative temperature variations. One may detect \textbf{half} periodicity close to the loop apex, where emission intensity modulation is relatively small. The line-of-sight projection affects the observation of Doppler shift significantly. A more accurate  estimate of the amplitude of velocity perturbation is obtained by de-projecting the Doppler shift by  a factor of $1-2\theta/\pi$ rather than the traditionally used $\cos\theta$. \textbf{If a loop is heated to the hotter wing, the intensity modulation could be overwhelmed by background emission, while the Doppler shift velocity could still be detected to a certain extent.}
\end{abstract}

\keywords{Sun: atmosphere --- Sun: corona --- Sun: oscillations --- magnetohydrodynamics (MHD) --- waves}

\section{Introduction}
\label{sec:intro}
Magnetohydrodynamic (MHD) waves are believed to play a significant role in the formation and dynamics of the solar atmosphere.
They may contribute significantly to coronal heating \citep[see reviews by][]{klimchuk2006,taroyan2009,parnell2012,arregui2015} and solar wind acceleration \citep[see e.g.,][]{ofman2010,vanderholst2014}. During the past decade, a number of MHD wave modes of coronal loops \textbf{were} detected with modern instruments, e.g., standing and propagating fast kink mode \citep{nakariakov1999,aschwanden1999,williams2002,vandoorsselaere2008}, fast sausage modes \citep{asai2001,melnikov2005}, standing and propagating slow mode \citep{wang2003a,wang2003b,demoortel2000,demoortel2002a,demoortel2002b,wang2009a,wang2009b,yuan2012sm,krishnaprasad2014}. 

MHD wave theory in structured plasma forms a solid basis for a wave-based plasma diagnostic technique -- MHD coronal seismology \citep[see][for recent reviews]{nakariakov2005,demoortel2012}. MHD seismology was successfully applied in estimating the coronal magnetic field \citep{nakariakov2001}, transverse loop structuring \citep{aschwanden2003}, \alfven transit times \citep{arregui2007}, polytropic index \citep{vandoorsselaere2011a}, thermal conduction coefficient \citep{vandoorsselaere2011a}, the magnetic topology of sunspots \citep{yuan2014cf,yuan2014lb, jess2013}, and the magnetic structure of large-scale streamers \citep{chen2010,chen2011}. \textbf{It could also be used to determine the coronal density scale height \citep{andries2005}, to quantify the expansion factor of the coronal loops \citep{verth2008}, and to probe the  characteristic spatial scale of randomly structured plasmas \citep{yuan2015rs}.}

The slow sausage mode was initially theorised by \citet{edwin1983}. It is a compressive mode characterised by axisymmetric longitudinal displacement of the plasma fluid. Gas pressure is the main restoring force. In a low-$\beta$ plasma, this mode does not cause significant contraction or expansion of the loop cross-section, nor a displacement of the loop axis. Standing slow modes are frequently observed as oscillations in the plasma emission intensity and Doppler shift velocity in hot flaring coronal loops \citep[$>6 \unit{MK}$, see review by][]{wang2011}. They are exclusively detected in the hot emission lines, i.e., \ion{Fe}{19} and  \ion{Fe}{21} lines as recorded by the Solar Ultraviolet Measurements of Emitted Radiation (SUMER) spectrograph onboard Solar and Heliospheric Observatory (SOHO) \citep{wang2002,wang2003a,wang2003b}, and \ion{S}{15} and \ion{Ca}{19} lines observed by the Bragg Crystal Spectrometer (BCS) onboard Yohkoh \citep{mariska2006}. Only recently, the Solar Dynamics Observatory (SDO)/Atmospheric Imaging Assembly (AIA) imager observed, in the 131\AA{} bandpass, multiple reflections of a propagating slow wave in a hot coronal loop \citep{kumar2013}. A slow compressive mode is found to be launched by repetitive magnetic reconnections occurring at one of the footpoints \citep{kumar2015}.  

Quasi-periodic pulsations (QPP) in solar and stellar flares are thought to be caused by MHD waves \citep{nakariakov2009,anfinogentov2013b}. Oscillations with periods at tens of minutes are ascribed to modulations by slow mode MHD waves \citep{vandoorsselaere2011b}, while short period (sub-minute) oscillations are suggested to be modulated by fast mode waves \citep{vandoorsselaere2011b,kupriyanova2013}. Recently, \citet{kim2012} detected oscillations during a M1.6 flare with the Nobeyama Radioheliograph (NoRH) and  the 335 \AA{} EUV channel of SDO/AIA. The intensities in radio and EUV emissions oscillates with a period of about 13.8 minutes and are damped with a decay time of about 25 minutes. \textbf{\citet{kim2012} interpreted the detected ten-minute time scale oscillations as standing slow magnetoacoustic waves. SUMER-like oscillations, but with shorter periodicities, are detected in soft X-ray emissions \citep{ning2014}.}

The standing slow mode wave oscillates with a period of about ten minutes and with a velocity amplitude of a few tens of kilometers per second \citep{wang2003b,wang2011}. Significant oscillation is normally detected at the loop apex and becomes absent at the footpoints \citep{wang2007}. This is consistent with a scenario that an anti-node in the density perturbation of the fundamental standing slow mode is located at the loop apex \citep{wang2007}. This kind of waves are damped within a few oscillation cycles, which are believed to be caused mainly by thermal conduction in hot flaring loops \citep{ofman2002,selwa2005}. 

\textbf{The fundamental slow standing mode of a hot loop appears to be triggered by asymmetric heating at one footpoint and is rapidly established within one wave cycle \citep{wang2005}. Pressure pulses launched close to a footpoint cannot excite a fundamental slow mode fast enough to compensate the strong damping \citep{selwa2005}. \citet{taroyan2005} showed analytically that the fundamental slow mode of a $6\unit{MK}$-loop could rapidly be excited by a single impulsive heating with a time scale that matches the loop period. Forward modelling using simple one dimensional hydrodynamic model were performed to distinguish the propagating and standing slow waves in both cool and hot loops \citep{taroyan2007,taroyan2008}. They also reported that Doppler shift variation is a more reliable observable to detect a slow wave, while intensity modulation would be phase-shifted by heating \citep{taroyan2007} or contaminated by the background plasma emissions \citep{taroyan2008}. 
}

MHD seismology with the standing slow mode was applied successfully to estimate the magnetic field strength and the time-dependent plasma temperature of a coronal loop \citep{wang2007}. The application of MHD seismology relies on several nontrivial factors, e.g., the analytical model, mode identification, line-of-sight (LOS) effect, and plasma emission. Forward modelling was attempted to help interpreting the observations correctly. \citet{cooper2003a,cooper2003b} investigated the LOS effect on imaging observation of the emission intensity variation of fast kink and sausage mode, and explained the intensity perturbations of fast wave trains observed by the Solar Eclipse Coronal Imaging System (SECIS) instrument \citep{williams2001,williams2002}. \citet{gruszecki2012} performed a three dimensional numerical simulation of the fast sausage mode of a plasma cylinder and investigated the geometric effect of simple LOS integration. \citet{antolin2013} and \citet{antolin2014} developed advanced models and included atomic emission effects, using the CHIANTI atomic database \citep{dere1997}. They found that the LOS effect and spatial resolution could significantly affect the intensity modulation and spectral characteristics of the fast sausage mode. \citet{reznikova2014,reznikova2015} used the same model and further investigated the gyrosynchrotron emission intensity variation using the Fast Gyrosynchrotron Codes \citep{fleishman2010}. The radio emission intensity of the fast sausage mode oscillate in phase for all frequencies, while for certain LOS angles, the optically thick and thin radio emission are anti-phase. It \textbf{opposes} previous findings of \citet{mossessian2012}, which did not consider the inhomogeneity of the emitting source along the LOS. \textbf{\citet{kuznetsov2015} used a semi-torus model to forward model the gyrosynchrotron radio emission of both propagating and standing slow modes in a curved magnetic structure.}

In this study, we perform forward modelling of standing slow modes of hot flaring coronal loops and predict \textbf{their} spectroscopic and imaging observational signatures. We use the slow wave model in hot coronal loops ($>6\unit{MK}$, see \secref{sec:model}) and utilize the CHIANTI v7.1 atomic database \citep{dere1997,landi2013} to synthesize plasma emission in the SUMER \ion{Fe}{19} line and SDO/AIA 094 \AA{} bandpass (see \secref{sec:method}). Then the results and conclusions are summarised in \secref{sec:result} and \secref{sec:disc}, respectively. 

\section{Model}
\label{sec:model}
\subsection{Standing slow mode}
In this study, we only consider a standing slow mode in a simple plasma cylinder \textbf{embedded in a uniform plasma. The magnetic field is parallel to the axis of the plasma cylinder (i.e., $z$-axis), $\bvec{B_0}=B_0 \bvec{\hat{z}}$. The equilibrium magnetic field $B_0$, plasma density $\rho_0$ and temperature $T_0$ are the piecewise functions of the  $r$-axis:
\begin{equation}
 B_0,\rho_0,T_0=\left\{
     \begin{array}{lr}
        B_\i,\rho_\i,T_\i &: r \leq a \\
        B_\e,\rho_\e,T_\e  &: r>a,
     \end{array}
   \right.
\end{equation}
where $a$ is the radius of the loop. Hereafter, we use subscript \singlequote{$\i$} and \singlequote{$\e$} to differentiate the internal and external equilibrium values of the loop system. 
} 
Effects of plasma stratification and loop curvature are ignored. We focus on observational features caused by optically thin plasma emission, LOS integration and instrument response function. We limit our study to the axisymmetric mode $m=0$ \citep[sausage mode, see][]{edwin1983}. 

The linearised ideal MHD equations \citep[see, e.g.,][]{ruderman2009} give the perturbed quantities on top of the magnetostatic equilibrium: 
\begin{align}
 \rho_1 &=-\div(\rho_0\bvec{\xi}), \label{eq:cont} \\
 \rho_0\frac{\partial^2\bvec{\xi}}{\partial t^2}&=-\grad{P_{\T1}} + \frac{1}{\mu_0}[ (\bvec{B_0}\cdot\grad)\bvec{b_1}+(\bvec{b_1}\cdot\grad)\bvec{B_0}],\label{eq:moment}\\ 
 \bvec{b_1}&=\curl(\bvec{\xi}\times\bvec{B_0}), \label{eq:mag}\\
 p_1-C_\s^2\rho_1&=\bvec{\xi}\cdot(C_\s^2\grad{\rho_0}-\grad{p_0})  \label{eq:entropy},
\end{align}
where $\bvec{\xi}$ is the Lagrangian displacement vector, $p_0$ is the equilibrium plasma pressure, $\rho_1$, $p_1$ and $\bvec{b_1}$ are the perturbed plasma density, pressure and magnetic field, $P_{\T1}=p_1+\bvec{b_1}\cdot\bvec{B_0}/\mu_0$ is the perturbed total pressure, $\mu_0$ is the magnetic permeability in free space. We define the key \textbf{characteristic speeds} to describe the loop system, $C_\s=\sqrt{\gamma p_0/\rho_0}$, $C_\A=B_0/\sqrt{\mu_0\rho_0}$, $C_\T=C_\A C_\s/\sqrt{C_\A^2+C_\s^2}$ are the acoustic, \alfven, and tube speed, respectively \citep{edwin1983}; $\omega_\s=C_\s k$, $\omega_\A=C_\A k$, $\omega_\T=C_\T k$ are the corresponding acoustic, \alfven, and tube frequencies, respectively, 
where $k=\pi n/L_0$ is the longitudinal wavenumber, $n$ is the longitudinal mode number, $L_0$ is the length of the loop, $\gamma=5/3$ is the adiabatic index. 

\eqref{eq:cont}-\ref{eq:entropy} are solved in cylindrical coordinates ($r,\phi,z$) with the boundary condition at $r=a$ where the radial displacement $\xi_r$ and the total pressure are kept in balance. In case of the standing slow mode with $m=0$, we Fourier-analyse the perturbed quantities by assuming $P_{\T1}=A\RR(r) \cos (\omega t) \cos (k z)$, where $A$ is the amplitude of the perturbed total pressure, and $\RR$ is defined as
\begin{equation}
\RR=\frac{\omega^2}{\rho_0(C_\s^2+C_\A^2)(\omega^2-\omega_\T^2)}\frac{\d}{\d t}\RR'.
\end{equation}
Here $\RR'$ is the relevant \singlequote{$\RR$} in \citet{antolin2013}. We use a $\cos (k z)$ longitudinal profile in total pressure perturbation, so the density (temperature) nodes are fixed at footpoints, while the longitudinal velocity perturbation follows a profile of $\sin (k z)$ and thus has a node at the loop apex for the fundamental mode. 
A key derivation is that 
\begin{equation}
\div{\bvec{\xi}}=-\frac{\omega^2 P_{\T1}}{\rho_0(C_\s^2+C_\A^2)(\omega^2-\omega_\T^2)}.
\end{equation}
The perturbed total pressure must satisfy 
\begin{equation}
\frac{\d^2P_{\T1}}{\d r^2}+\frac{\d P_{\T1}}{r\d r}-\kappa_r^2P_{\T1}=0, \label{eq:ptr}
\end{equation} 
where $\kappa_r^2=\frac{(\omega_\s^2-\omega^2)(\omega_\A^2-\omega^2)}{(\omega_\s^2+\omega_\A^2)(\omega_\T^2-\omega^2)}k^2$ is a modified radial wavenumber and has the dimensionality of wavenumber $k$. \eqref{eq:ptr} holds for both internal and external plasmas, where all quantities are piecewise functions of $r$ \textbf{(\figref{fig:loop1})}. \eqref{eq:ptr} gives 
\begin{equation}
\RR=\left\{
     \begin{array}{lr}
       J_0(|\kappa_{r\i}| r) & : r \leqslant a \\
       K_0(\kappa_{r\e}r) & : r>a.
     \end{array}
   \right.
\end{equation}
By matching the boundary conditions, we obtain the dispersion relation for the fast and slow sausage body mode (\citet{edwin1983}, $\kappa_{r\e}^2>0$ and $\kappa_{r\i}^2<0$, hence we re-define $|\kappa_{r\i}|=\sqrt{-\kappa_{r\i}^2}$)
\begin{equation}
\frac{\kappa_{r\e}}{\rho_\e(\omega_{\A\e}^2-\omega^2)}\frac{K'_0(\kappa_{r\e} a)}{K_0(\kappa_{r\e}a)}=\frac{\kappa_{r\i}}{\rho_\i(\omega_{\A\i}^2-\omega^2)}\frac{J'_0(|\kappa_{r\i}| a)}{J_0(|\kappa_{r\i}| a)}, \label{eq:disp}
\end{equation}
where $J_0$ is the 0-th order Bessel function of the first kind and $K_0$ is the 0-th order modified Bessel functions of the second kind; $J'_0$ and $K'_0$ are the corresponding derivatives with respect to $\kappa_{r}r$. The perturbed thermodynamic quantities, \textbf{which affect plasma emissions}, are the velocity vector $\bvec{v}=\partial\bvec{\xi}/\partial t$, the plasma density $\rho$, and the temperature $T$:
\begin{align}
 v_r &=-\frac{A\d \RR/ \d r \omega}{\rho_0(\omega^2-\omega_\A^2)}\sin (\omega t) \cos (k z), \label{eq:vr} \\
 v_z &=-\frac{A\RR C_\T^2 k \omega}{\rho_0V_\A^2(\omega^2-\omega_\T^2)}\sin (\omega t) \sin (k z),\\
  v_\phi& =0, \\
 \rho_1&=\frac{A\RR }{(C_\s^2+C_\A^2)}\frac{\omega^2}{(\omega^2-\omega_\T^2)} \cos (\omega t) \cos (k z), \\
 T_1&=\frac{A\RR (\gamma-1)T_0}{\rho_0(C_\s^2+C_\A^2)}\frac{\omega^2}{(\omega^2-\omega_\T^2)} \cos (\omega t) \cos (k z), \label{eq:temp}
\end{align}
where $T_0$ is the equilibrium plasma temperature. The thermodynamic variables are related by the equation of state for fully ionized hydrogen $p=2k_\mathrm{B}\rho T/m_\mathrm{p}$, where $\rho=\rho_0+\rho_1$, $T=T_0+T_1$, $k_\mathrm{B}$ is the Boltzmann constant, $m_\mathrm{p}$ is \textbf{proton mass}.

\eqref{eq:ptr}-\ref{eq:temp} holds for both the fast and slow sausage mode. The solution depends on whether the equations are solved in the \alfven $[\omega_{\A\i}, \omega_{\A\e}]$ or acoustic $[\omega_{\T\i},\omega_{\s\i}]$ \textbf{frequency range} \citep{sakurai1991}. We note that $v_r/v_z\propto(\omega^2-\omega_\T^2)/(\omega^2-\omega_\A^2)$. Therefore, in the \alfven \textbf{frequency range}, $v_r/v_z\gg1$, it corresponds to the fast sausage mode \citep{antolin2013,reznikova2014}; while in the acoustic \textbf{frequency range}, $v_r/v_z\ll1$, this corresponds to the slow sausage mode \citep[see, e.g.,][]{moreels2013}, which is the topic of this study.  

\subsection{Hot flaring coronal loop}
\label{sec:model_loop}

\textbf{Hot coronal loops are complex and highly dynamic structures heated to 2-30 MK by flares \citep[see a review by][]{reale2014}. A coronal loop may have unresolved fine structures \citep{priest2002,vandoorsselaere2014} or multiple-strands \citep{peter2013,antolin2012,antolin2014,antolin2015,schullion2014}. Heating/cooling \citep[see e.g.,][]{klimchuk2006,hood2009} and the associated flows \citep[see][]{winebarger2002} are usually detected at the footpoints of the loops. MHD wave theory \citep[see e.g.,][]{edwin1983,sakurai1991,goossens2011} normally assumes that a quiescent loop is in equilibrium with the ambient plasma, therefore the heating and cooling time scale should be sufficiently longer than the MHD time scales (in order of minutes in the case of slow modes). If a loop is subject to active heating or cooling, the Wentzel-Kramers-Brillouin (WKB) approximation can be used (similar to \citet{ruderman2011b,ruderman2011a}).} 

\textbf{In this study, we are only concerned with the observational features of established standing slow modes, even if the WKB approximation is violated, i.e., the MHD slow mode time scale is in the same order of the heating or cooling time scale, the result could still be used to identify a standing slow wave based on stepwisely-defined quasi-equilibriums.} We set up a hot and dense flaring coronal loop with typical parameters that are observed by SUMER \citep{wang2011}. Our loop measures $L_0=100\unit{Mm}$  in length and $a=5\unit{Mm}$ in radius. The loop is filled with plasma \textbf{with} a density $\rho_0=1.4\cdot10^{-11}\unit{kg\, m^{-3}}$ (the electron density $n_{\e0}=8.5\cdot10^{9}\unit{cm^{-3}}$) and a temperature of $T_\i=6.4\unit{MK}$. The internal magnetic field is fixed at $B_\i=40\unit{G}$. We choose a density ratio of $\rho_\i/\rho_\e=5$, a temperature ratio $T_\i/T_\e=1.5$, and a magnetic field strength ratio $B_\i/B_\e=0.91$. The ratios of the plasma parameters are set in the typical range of flaring coronal loops. Changes in these ratios will not significantly affect the result, since in a slow mode the longitudinal perturbations are strictly confined within the plasma cylinder \textbf{and are more than four orders of magnitude stronger than the perturbations to the ambient plasma}. In this setup, the internal and external plasma are typical coronal fluid with plasma beta of $\beta_\i=0.23$ and $\beta_\e=0.02.6$, respectively. The acoustic speeds are $C_{\s\i}=420\unit{km\, s^{-1}}$, $C_{\s\e}=340\unit{km\, s^{-1}}$, while the \alfven speed are $C_{\A\i}=950\unit{km\, s^{-1}}$ and $C_{\A\e}=2300\unit{km\, s^{-1}}$ \textbf{(\tabref{tab:loop})}. These speeds are typical values observed in the solar corona \citep[see, e.g.,][]{aschwanden2005}. \textbf{We also investigate the slow modes in hot loops at $T_i=8.8,12, 15,20\unit{MK}$ (see \tabref{tab:loop}), and explore how the properties would deviate from the case of the $6.4\unit{MK}$ loop. The total pressure perturbation is kept unchanged, the amplitude of perturbed density, temperature and velocity will vary according to the equillibrium temperature (\tabref{tab:loop}). However, the ampltiude of the perturbed quantities will not affect overall result, since in linearized MHD wave modes they are scalable. In the following text, we refer to the case of slow wave in the $6.4\unit{MK}$ loop by default, and other cases are specified as otherwise in \secref{sec:temp}. 
}

For $n=1$ (the fundamental longitudinal mode), $ka=0.157$ is in the long wavelength limit. We use $A_\i=0.01$ so that the velocity perturbation is about $57\unit{km\, s^{-1}}$ and the density perturbation is about $12\%$ of the equilibrium value. For $n=2(3)$, we use $A_\i=0.02(0.04)$. The velocity and density perturbation are about $41(48)\unit{km\, s^{-1}}$ and $9(11)\%$ of the equilibrium value, respectively \textbf{(\tabref{tab:loop})}. These amplitude values are chosen to agree with the observed Doppler shift velocities in \citet{wang2011}.  By solving the dispersion relation \eqref{eq:disp}, we obtained the periods $P_0=520 \unit{s},260\unit{s},170\unit{s}$ for the $n=1,2,3$ modes,  respectively. 

\begin{figure}[ht]
\centering
\includegraphics[width=0.5\textwidth]{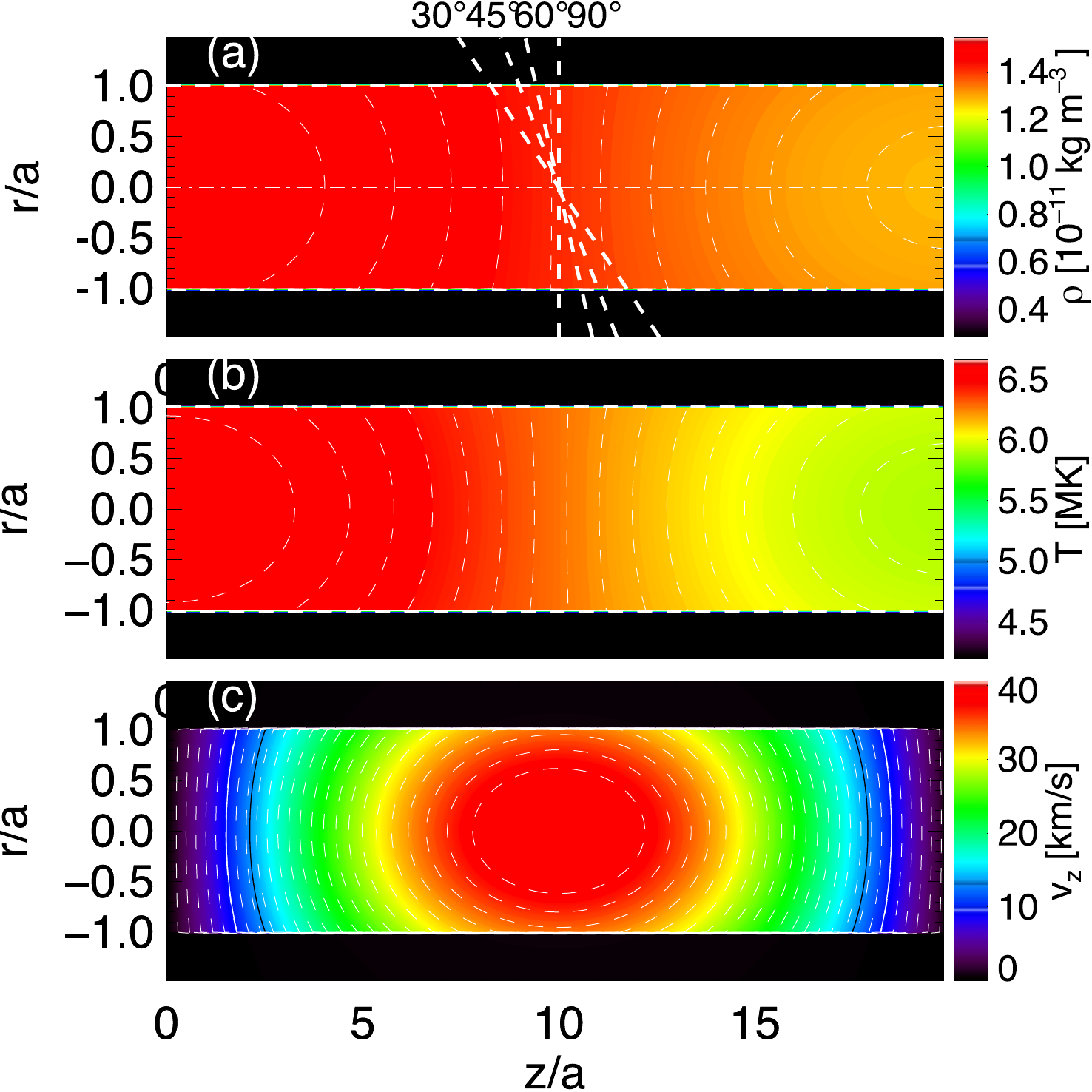}
\caption{A snapshot of $\rho(r,z)$ (a), $T(r,z)$ (b), and $v_z(r,z)$ (c) at $t=P_0/8$. The dashed lines indicate the position of the rays at LOS of $30\deg$, $45\deg$, $60\deg$, and $90\deg$, respectively.}
\label{fig:loop1}
\end{figure}

Starting from the equilibrium loop model, we construct a discrete standing slow wave model as specified by \eqref{eq:vr}-\ref{eq:temp}. The simulation domain ranges from $[0,1.5a]$, $[0,2\pi]$, $[0,L_0]$ for $r$-, $\phi$-, and $z$-axes, with grid cells of  $150\times180\times300$, respectively. \figref{fig:loop1} illustrates a snapshot of the $\rho$, $T$ and $v_z$ distribution at $t=P_0/8$ for the $n=1$ mode. An anti-node in terms of density perturbation is present at $z=L_0/2$, this is in agreement with SUMER observations that Doppler shift oscillations are usually effectively detected at the loop apex rather than the loop footpoints \citep{wang2007}. The density and temperature perturbations are in phase and are a quarter period out of phase, with respect to both time and space, with the longitudinal velocity. \textbf{We note that the density/temperature and velocity perturbations are in phase for propagating slow waves \citep{sakurai2002,wang2009b}, therefore, a mix of propagating and standing wave would lead to the detection of a rather complex phase lag \citep{wang2009a}. In this study, we focus on purely standing slow MHD waves to obtain guidelines for observations.}

\begin{table*}
\begin{center}
\caption{Parameters of the loop systems and the standing slow modes \label{tab:loop}}
\begin{tabular}{lrrrrr}
\tableline\tableline
Loops & $T_\i=6.4\unit{MK}$ &  $T_\i=8.8\unit{MK}$ & $T_\i=12\unit{MK}$ & $T_\i=15\unit{MK}$ & $T_\i=20\unit{MK}$  \\
\tableline
 $L_0 \unit{[Mm]}$  & 100   & 100   & 100  & 100  & 100 \\
 $a\unit{[Mm]}$     & 5.0   &  5.0  & 5.0  & 5.0  & 5.0 \\
 $B_\i\unit{[G]}$   & 40.0  &  40.0 & 40.0 & 40.0 & 40.0  \\
 $B_\e\unit{[G]}$   & 43.8  &  45.3 & 47.0 & 48.7 & 51.6 \\
 $\rho_\i\unit{[10^{-11}kg\, m^{-3}]}$ & $1.4$ &  $1.4$ & $1.4$ & $1.4$&  $1.4$ \\
 $n_{e\i}\unit{[10^{9}cm^{-3}]}$ & $8.5$ &  $8.5$&  $8.5$ &  $8.5$&  $8.5$ \\
 $\rho_\i/\rho_\e$  & 5.0   & 5.0 & 5.0 & 5.0  & 5.0\\
 $T_\i\unit{[10^6K]}$ & $6.4$  & $8.8$ & $12.0$  & $15.0$ & $20.0$ \\
 $T_\i/T_\e$ & 1.5 & 1.5  &1.5  &1.5 & 2.0 \\
 $\beta_i$ & 0.23 & 0.32 & 0.44 & 0.55 & 0.74\\
 $\beta_e$ & 0.026 & 0.034& 0.043 &0.050 & 0.044 \\
 $C_{\s\i}\unit{[km\,s^{-1}]}$ &  $420$ & $490$  & $570$  & $640$ & $740$ \\
 $V_{\A\i}\unit{[km\, s^{-1}]}$ & $950$ & $950$  & $950$  & $950$ & $950$\\
 $C_{\T\i}\unit{[km\, s^{-1}]}$ & $380$ & $440$  & $490$  & $530$ & $580$\\
 $C_{\s\e}\unit{[km\, s^{-1}]}$ & $340$ & $400$  & $470$  & $520$ & $530$ \\
 $V_{\A\e}\unit{[km\, s^{-1}]}$ & $2300$& $2400$ & $2500$ & $2600$& $2700$\\
 $C_{\T\e}\unit{[km\, s^{-1}]}$ & $340$ & $400$  & $460$  & $510$ & $520$\\
                               & \multicolumn{5}{c}{$n=1,A=0.01$}   \\
 $v_z^0\unit{[km\, s^{-1}]}$ \tablenotemark{a} &  $57$ & $47$ & $38$ & $31$  & $28$ \\
 $\rho_1^0/\rho_\i$                           & 0.12 & 0.08 & 0.06 & 0.04 &  0.03 \\
 $T_1^0/T_\i$                                 & 0.08& 0.06& 0.04 & 0.03 & 0.02 \\
 			      & \multicolumn{5}{c}{$n=2,A=0.02$}  \\
 $v_z^0\unit{[km\, s^{-1}]}$  &  41 & \nodata & \nodata & \nodata & \nodata \\
 $\rho_1^0/\rho_\i$& 0.09 & \nodata &  \nodata& \nodata & \nodata\\
 $T_1^0/T_\i$& 0.06&\nodata   &\nodata  & \nodata& \nodata \\
 		& \multicolumn{5}{c}{$n=3,A=0.04$ }  \\
 $v_z^0\unit{[km\, s^{-1}]}$  &  48 & \nodata& \nodata  &\nodata & \nodata   \\
 $\rho_1^0/\rho_\i$ & 0.11 & \nodata & \nodata & \nodata & \nodata \\
 $T_1^0/T_\i$ & 0.07 &\nodata  & \nodata & \nodata & \nodata \\
\tableline
\end{tabular}
\tablenotetext{a}{The superscript $0$ here and thereafter denotes the amplitude of the perturbed quantities in  \eqref{eq:vr}-\ref{eq:temp} excluding the spatial and temporal terms.}
\end{center}
\end{table*}

\section{Methods}
\label{sec:method}
\subsection{Spectroscopic Modelling}
\label{sec:method_spect}
\begin{figure}[ht]
\centering
\includegraphics[width=0.5\textwidth]{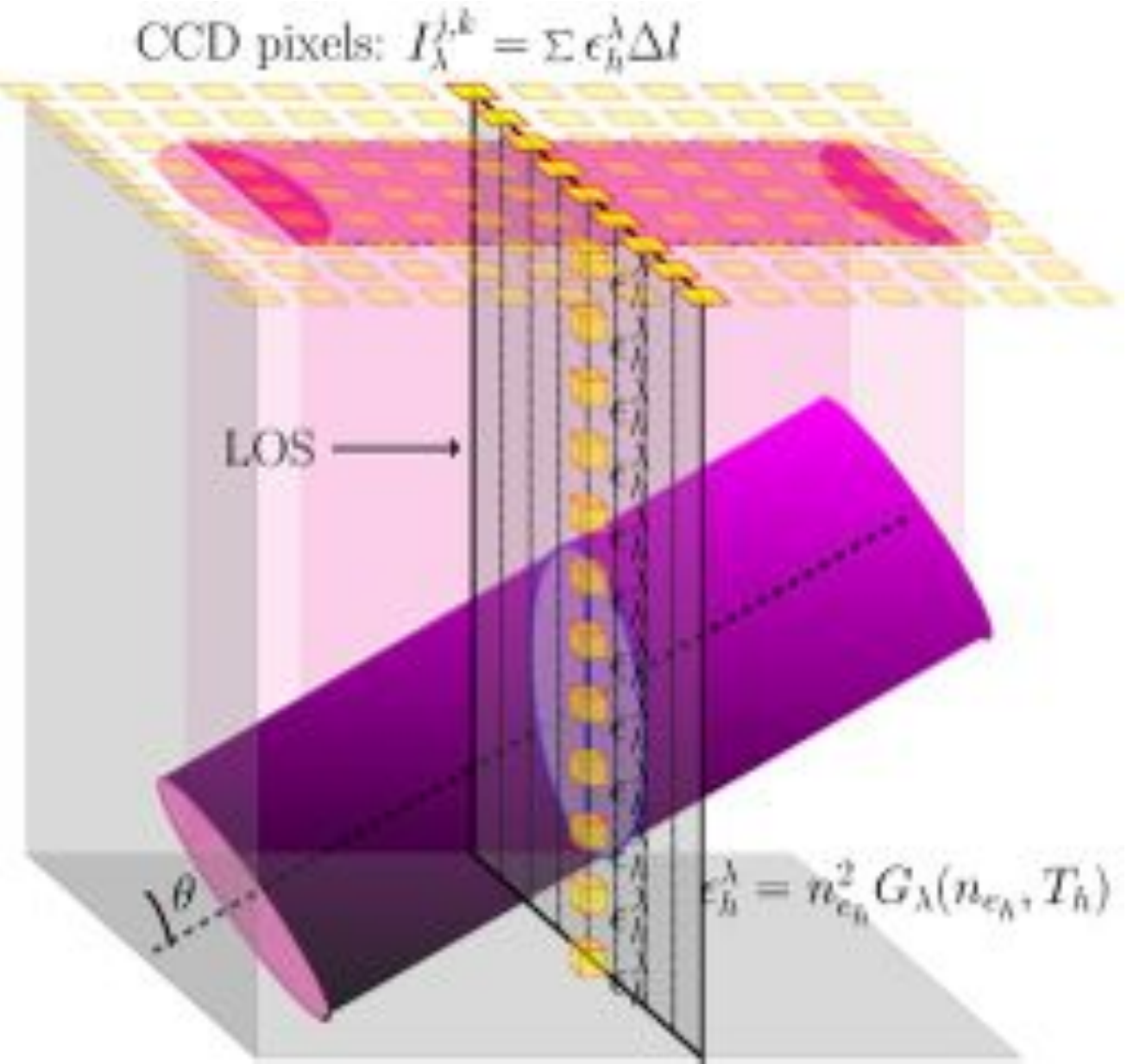}
\caption{\textbf{Schematic diagram illustrates the forward modelling method and LOS integrations.}\label{fig:fomo}}
\end{figure}

We are concerned with synthesizing the extreme ultraviolet (EUV) emission intensity $I_{\lambda_0}\unit{[ergs\,cm^{-2}\,s^{-1}\,sr^{-1}]}$ of a specific spectral line $\lambda_0$ for optically thin plasma along the LOS \citep{dere1997},
\begin{equation}
I_{\lambda_0}=\frac{A_b}{4\pi}\int{G_{\lambda_0}(n_e,T)n_e^2\d l}, \label{eq:int}
\end{equation}
where $A_b$ is the abundance of the emitting element relative to hydrogen, $G_{\lambda_0}\unit{[ergs\,cm^{3}\,s^{-1}]}$ is the contribution function that contains the terms relative to atomic physics, $\d l$ is an infinitesimal element length along the LOS. 

To calculate the integrated emission intensity in \eqref{eq:int}, we generate a look-up table for the \ion{Fe}{19} \feline line using the CHIANTI v7.1 atomic database \citep{dere1997,landi2013}. The look-up table is sampled at a uniform mesh of size $200\times200$ grid points at $\log{n_\e} \unit{[cm^{-3}]}\in [8,11]$ and $\log{T}\unit{[K]}\in [6.2,7.7]$. We used the CHIANTI collisional ionization equilibrium file $\mathit{chianti\_v7.ioneq}$ and coronal abundance $\mathit{sun\_coronal\_2012\_schmelz.abund}$ \citep[also see][]{antolin2013}\footnote{The source of the forward modelling code (FoMo) is available at https://wiki.esat.kuleuven.be/FoMo}. The choices of ionisation and abundance files do not affect our results at all, since we are only concerned with the relative intensity perturbation and Doppler shift caused by the MHD waves.

\textbf{\figref{fig:fomo} illustrates the forward modelling method and how the LOS integration (\eqref{eq:int}) is implemented numerically.} For each grid point, the emissivity $\epsilon^{\lambda_0}=G_{\lambda_0}n_e^2$ is calculated and is spread to a Gaussian spectrum with the width determined by the thermal broadening of the spectral line. The spectrum covered in this study is centered at \feline with a range of $\pm0.5\unit{\AA{}}$ ($\pm130\unit{km/s}$). This is a \ion{Fe}{19} EUV line, in which most spectroscopic observations on standing slow wave were performed \citep[see][]{wang2011}. We used 60 wavelength values to sample the spectrum. It corresponds to a spectral resolution of $\Delta\lambda=16.7\unit{m\AA{}}$ or $\Delta v_\D=4.5\unit{km/s}$, which is sufficient to resolve the spectrum. Then along the LOS, the emission wavelength of the elementary plasma fluid is modified by the velocity perturbation caused by the wave and is re-binned into the discrete spectrum. By assuming a Gaussian distribution for the integrated spectrum, we obtain the emission intensity $I_{\lambda_0}$ and Doppler shift velocity $v_D$.

We perform the calculations for LOS angle $\theta=30\deg,45\deg,60\deg$, and $90\deg$, respectively. The projected plane-of-sky has a mesh grid of $N_x\times N_z=45\times300\sin\theta$, so that the synthesized LOS emission \textbf{plane has a pixel size of $0.33\times0.33\unit{Mm^2}$ (\figref{fig:fomo})}. 

\subsection{Imaging Modelling}
\label{sec:method_imager}

To synthesize the observational features of SDO/AIA channels, we calculated the AIA temperature response function $K_\alpha(n_\e,T) [\unit{DN\,cm^5\,s^{-1}}]$ for bandpass $\alpha$ \citep{boerner2012}:
\begin{equation}
K_\alpha(n_e,T)=\int_{0}^{\infty}G(\lambda, n_e,T) R_\alpha(\lambda)\d\lambda,\label{eq:resp}
\end{equation}
where $R_\alpha(\lambda)[\unit{DN\,cm^2\,sr\,photon^{-1}}]$ is the instrument-wavelength response function. This is the product of the solid angle occupied by a unit surface ($0.6\arcsec\times0.6\arcsec$) relative to the telescope and the value calculated with the $\mathit{aia\_get\_resp.pro}$. $G(\lambda, n_e,T)[\unit{photon\,cm^3\,s^{-1}\,sr^{-1}}]$ is the contribution function calculated with the $\mathit{isothermal.pro}$ routine in CHIANTI \citep[see][]{delzanna2011}. Then the flux $F_\alpha(\bvec{x})[\unit{DN\,s^{-1}}]$ at pixel $\bvec{x}$ is integrated along the LOS, 
\begin{equation}
F_\alpha(\bvec{x})=\int_{l}K_\alpha(n_\e,T)n_e^2\d l.\label{eq:aia}
\end{equation}
\textbf{\eqref{eq:aia} is solved using the same algorithm as \eqref{eq:int} ( \figref{fig:fomo}), however, the intensity is obtained by summing up the contributing emissions in all effective wavelengths (\eqref{eq:resp}) rather than spreading into a spectrum (see \secref{sec:method_spect}).}
A look-up table for each AIA bandpass is sampled at a uniform mesh of size $200\times200$ grid points at $\log{n_\e} \unit{[cm^{-3}]}\in [8,11]$ and $\log{T}\unit{[K]}\in [4,8]$. 

We synthesized the AIA 94\AA{} channel that would image our flaring loops ($>6.4\unit{MK}$). The resultant pixel size is kept uniform at $0.6\arcsec\times0.6\arcsec$, therefore we choose a mesh grid of $N_x\times N_z=35\times230\sin\theta$ for output and perform the calculations for LOS angles $\theta=30\deg,45\deg,60\deg$ and $90\deg$, respectively. Point spread functions \citep[PSF, see][]{poduval2013,antolin2013} would only have marginal effect at the edges and the cylinder boundaries. Moreover, we use long wavelength limits and the plasma motions are predominantly longitudinal, thus PSF effect is neglected in this study.  

\section{Results}
\label{sec:result}
In this section, we divide the results into two categories that observers mainly use to detect standing slow mode in a coronal loop. As we use linearization of the MHD wave model and perform geometric integration by modelling the atomic emissions, the results could be scaled to the range of interests. In the following subsections, we present a typical observation mode of a spectrograph or imager and study other effects that may affect the observations.
 
\subsection{Spectroscopic observations}
\label{sec:result_spect}
\subsubsection{Typical observation}
\begin{figure*}[ht]
\centering
\includegraphics[width=0.7\textwidth]{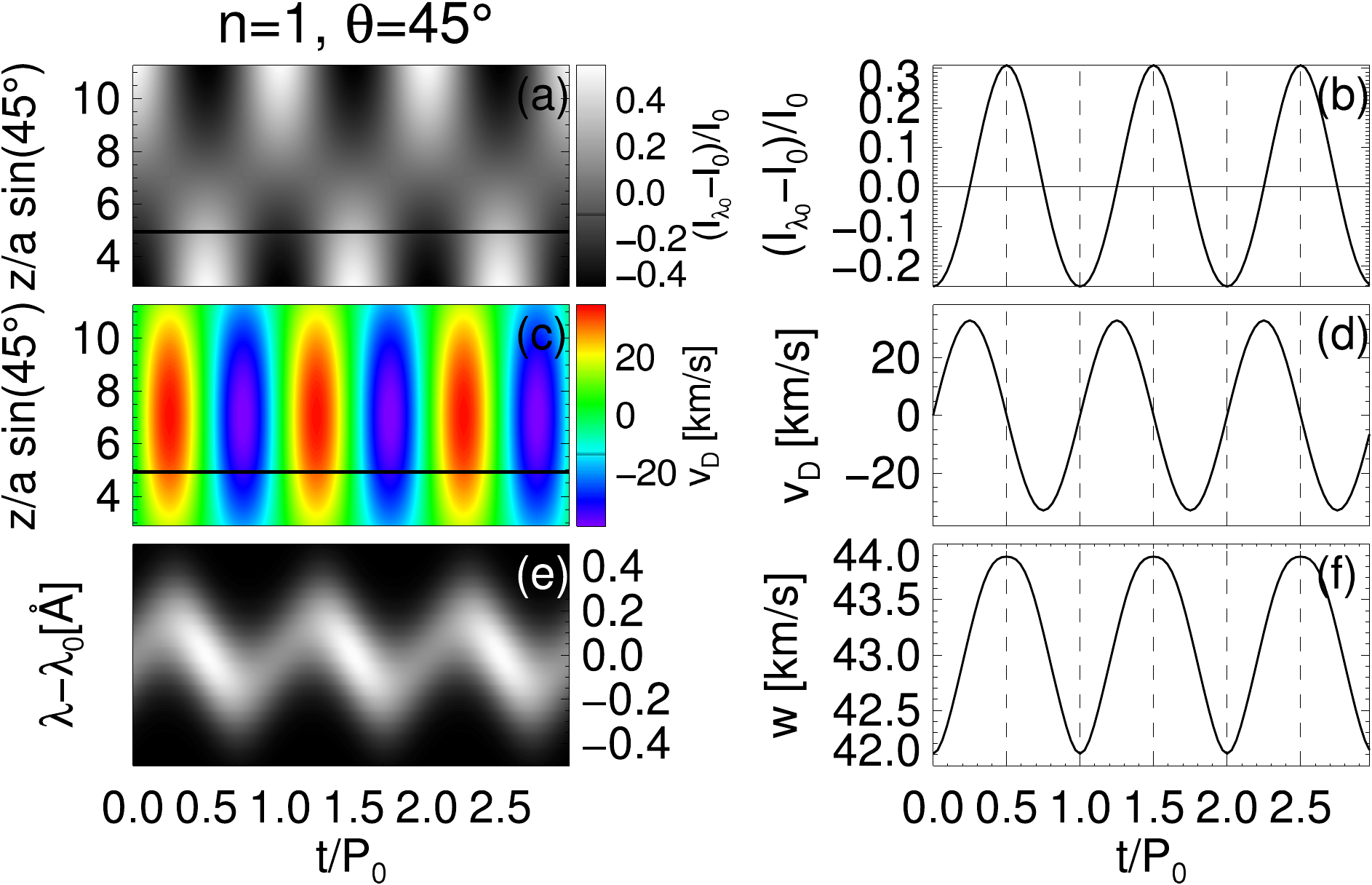}
\caption{(a) The baseline ratio time-distance plot of the relative intensity variation $(I_{\lambda_0}-I_0)/I_0$ along the central axis of the loop. (c) The time-distance plot of $v_z$ along the central axis of the loop. The solid line denotes the position where we took time series of $(I_{\lambda_0}-I_0)/I_0$ as shown in panel (b), (d) shows the variation of the Doppler shifts, (f) shows the variation of line widths and (e) shows the variation of the spectral line. All the information is extracted for mode $n=1$ at a viewing angle $\theta=45\deg$. }
\label{fig:spect1_45}
\end{figure*}

We first mimic a sit-and-stare campaign of a spectrograph, e.g., SUMER. The slit is ideally placed to fully cover the central line of our loop with a viewing angle of $45\deg$. \figref{fig:spect1_45} presents typical observables of a spectrograph. The intensity perturbation (\figref{fig:spect1_45}{a}) has a larger value at the loop footpoints than at the loop apex, while the Doppler shift $v_\D$ (\figref{fig:spect1_45}{c}) shows an opposite spatial pattern: it has a maximum at the loop apex. This is in agreement with the standing slow wave model (see \secref{sec:model_loop}), in which the longitudinal velocity and density are phase shifted in space by a quarter wavelength. Time series (\figref{fig:spect1_45}{b, d})  taken at a position off the loop apex and footpoints show a typical observation, which could be directly compared with Figure 2 in \citet{wang2011}: the intensity perturbation oscillates with a quarter-period out of phase in time with the Doppler shift $v_\D$. We also show that intensity variations are in phase with line width $w$ oscillations (\figref{fig:spect1_45}(b,f)). \figref{fig:spect1_45}{e} illustrates a typical spectrum that would be observed in a sit-and-stare mode: the spectral line is Doppler shifted by the standing slow wave and is also broadened by the perturbed velocity and temperature along LOS. Spectral observations at a slit position could provide the first signal of a MHD wave. The line width variation is not reported so far by any observations, its temporal variation and phase relations with other observables could be used to identify the wave mode. This may be due to the low amplitude of line width oscillation (about $1\unit{km/s}$), thus it is beyond the detection capability of current instruments. 

\begin{figure*}[ht]
\centering
\includegraphics[width=0.7\textwidth]{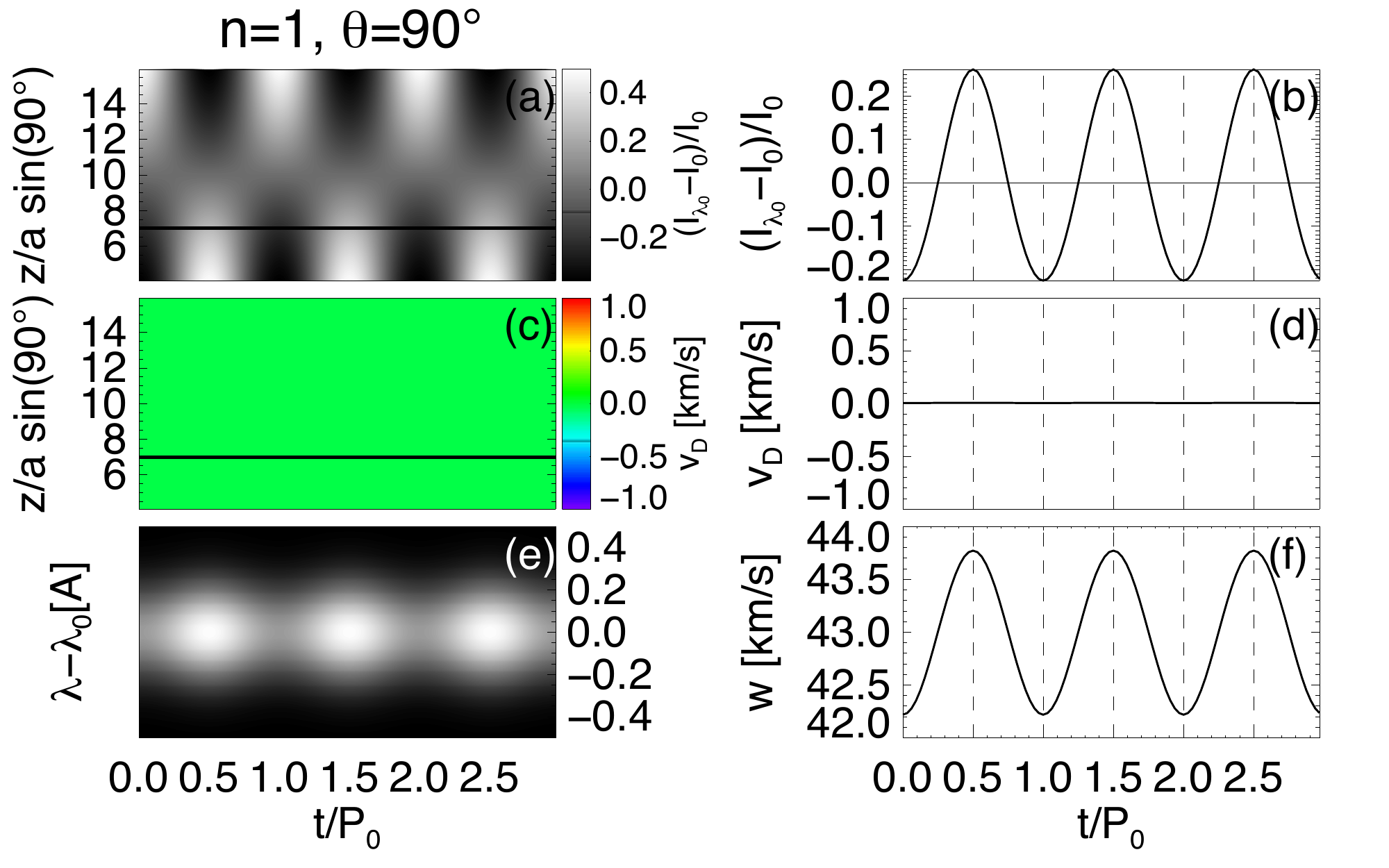}
\caption{The same as \figref{fig:spect1_45}, but for $n=1$, $\theta=90\deg$. }
\label{fig:spect1_90}
\end{figure*}

When the viewing angle is normal to the loop, i.e., $90\deg$ (\figref{fig:spect1_90}), the intensity and line width broadening modulation are still significant and are in phase with each other. However, the Doppler shift $v_D$ oscillation becomes below noise level and is not detectable. If a slit is placed over the loop apex (anti-node), then the intensity and line width modulation will be very small. 

A measure of goodness-of-fit $\chi^2$ to a Gaussian spectrum is not investigated in this study (see e.g., \citet{antolin2013}). It measures the level of goodness that a combination of multiple temperature spectral components could approximate a single Gaussian spectrum. For the standing slow mode, the plasma motions are predominately longitudinal, so plasma advections across the cylinder boundary are negligible. Therefore, the spectra rarely deviate from a Gaussian shape. In our study, the $\chi^2$ measures at the order of $10^{-4}$. It means the Gaussian profiles are in accord with the spectra or the error variance has been overestimated. The latter is true in our case, since we do not have error bars associated with the spectra and unity is used as error variance.  

\subsubsection{LOS effect}
\label{sec:los}
\begin{figure*}[ht]
\centering
\includegraphics[width=0.32\textwidth]{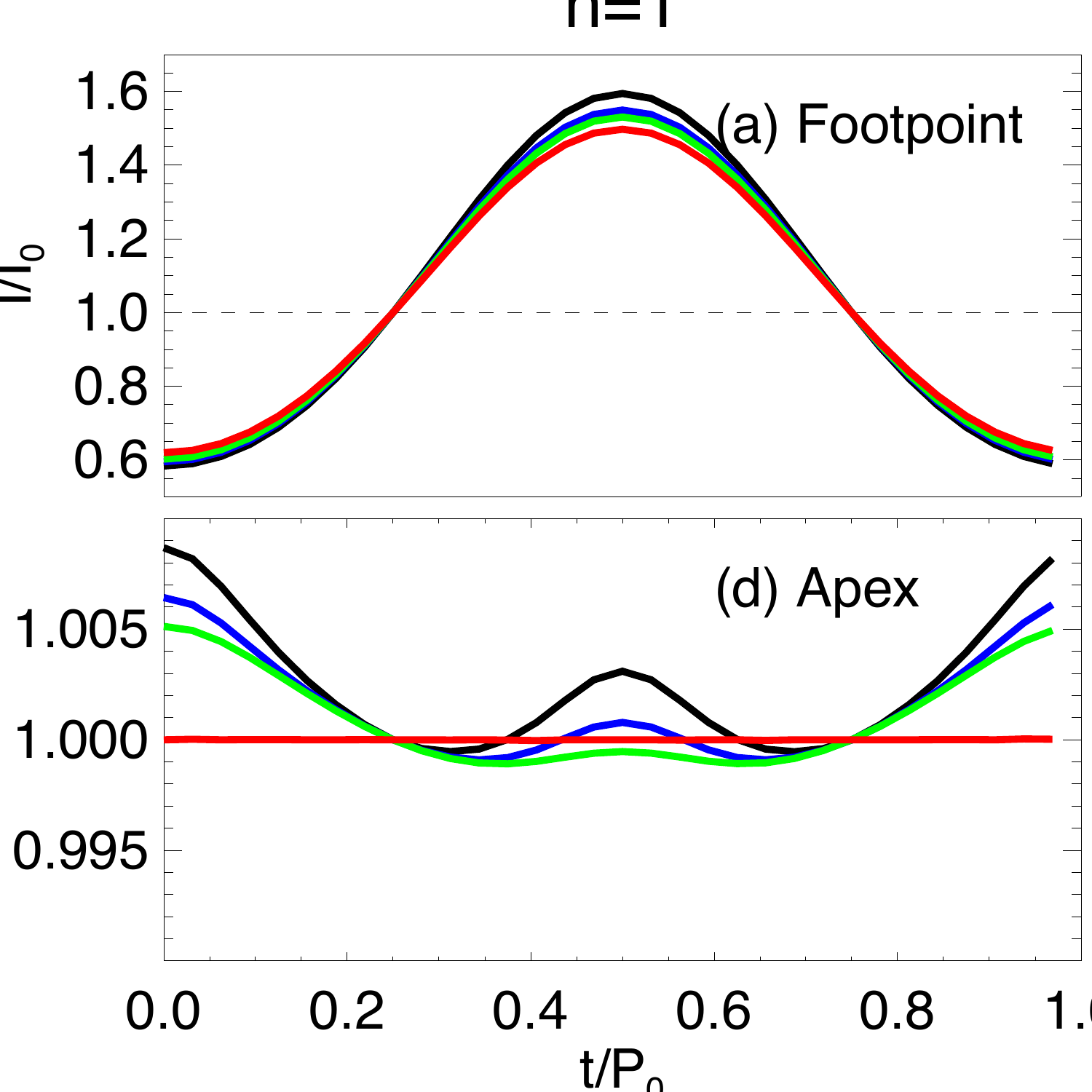}
\includegraphics[width=0.32\textwidth]{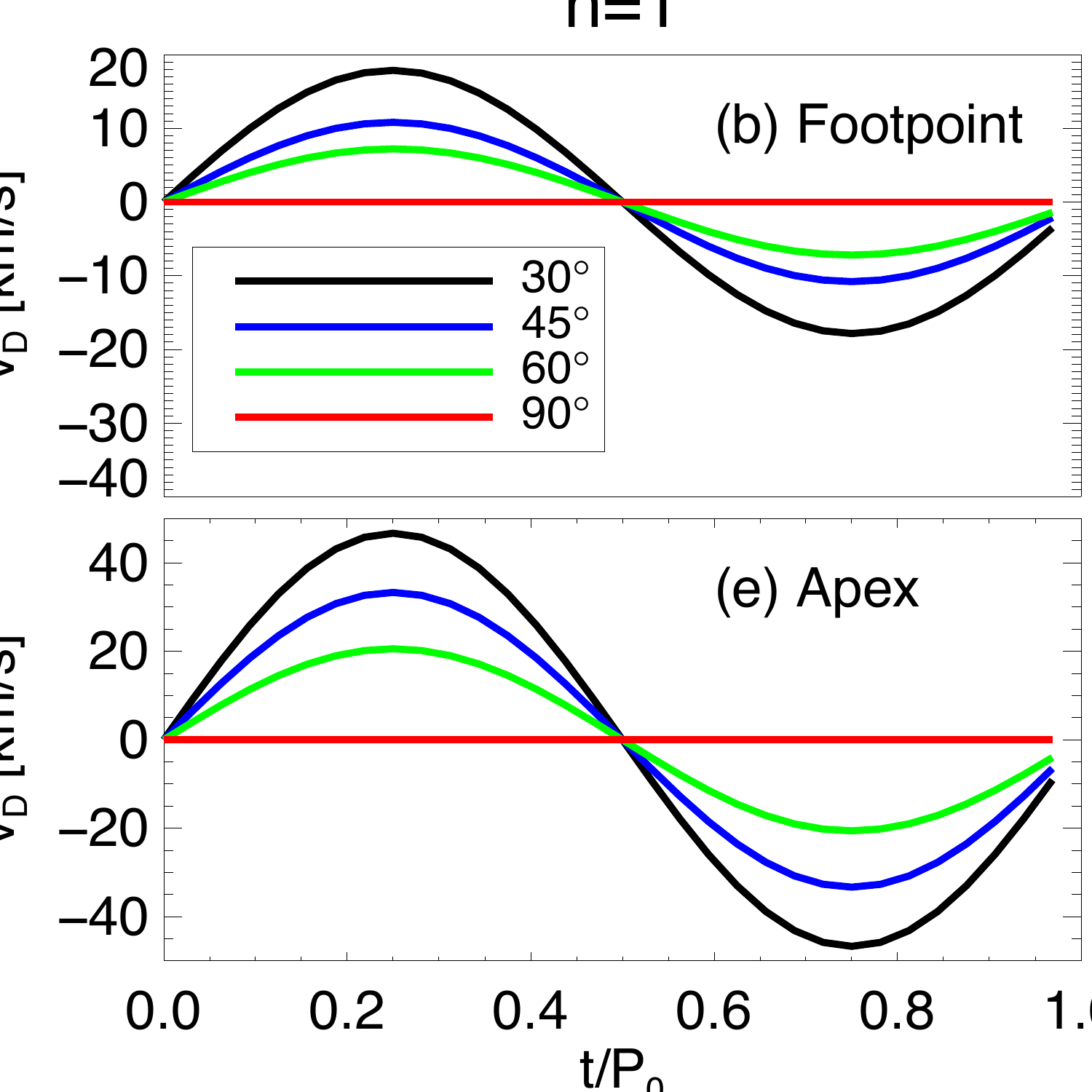}
\includegraphics[width=0.32\textwidth]{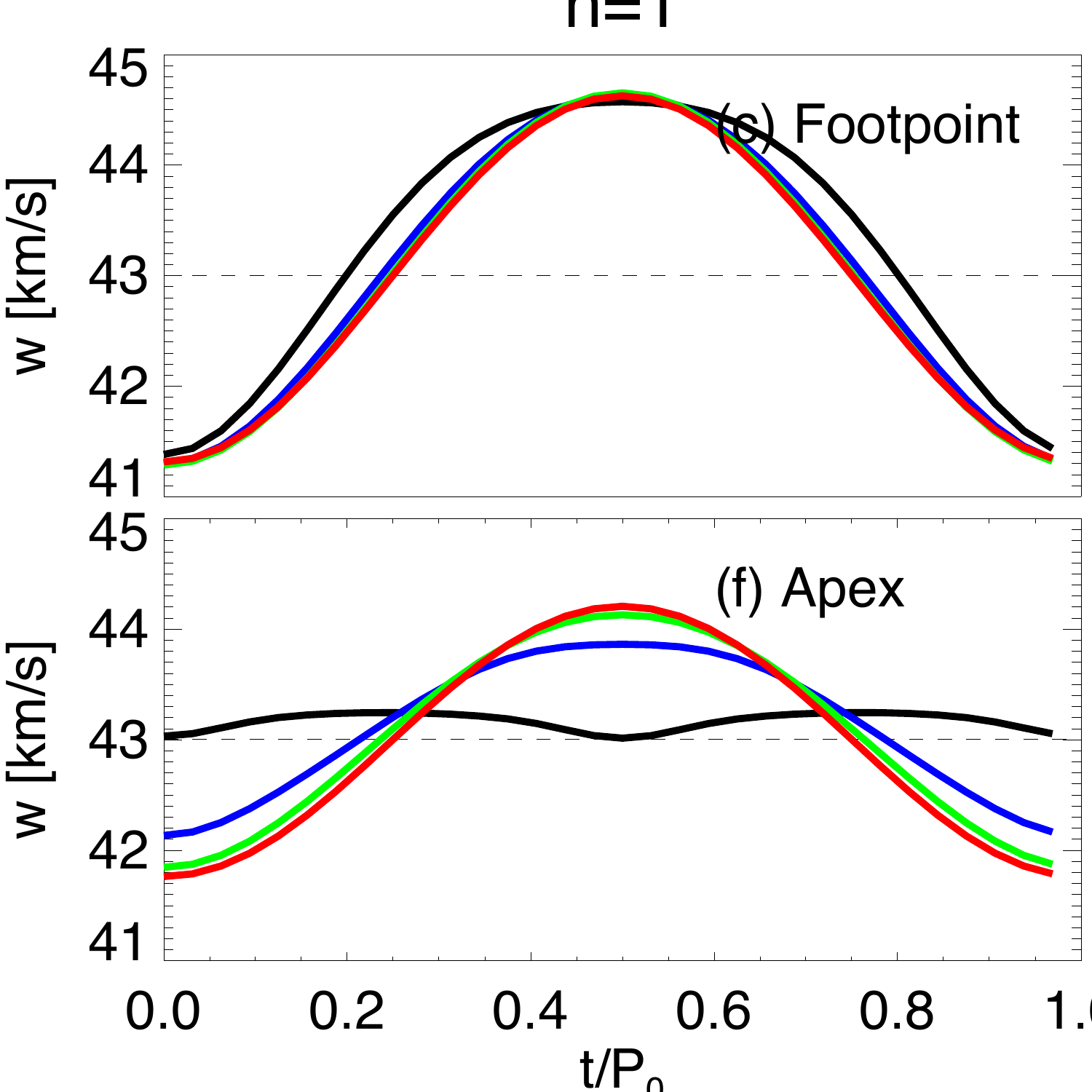}
\caption{One period variation of the normalized emission intensity $I/I_0$ (a, d), Doppler shift velocity $v_D$ (b, e), and line-width $w$ (c, f) observed at the loop footpoint and \textbf{apex} for the $n=1$ mode.}
\label{fig:spect_los1}
\end{figure*}

\begin{figure}[ht]
\centering
\includegraphics[width=0.5\textwidth]{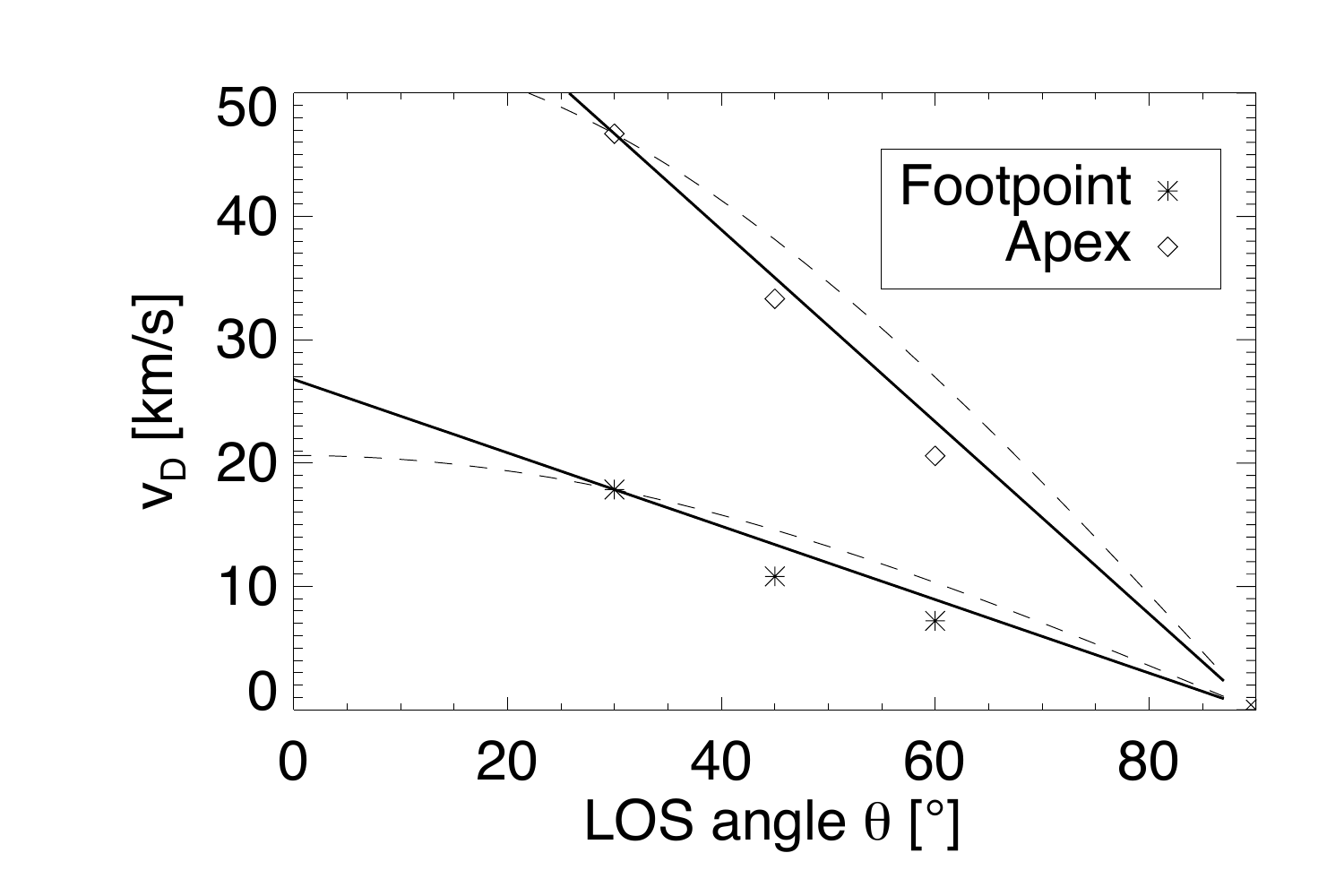}
\caption{The Doppler shift velocity amplitude as a function of LOS viewing angle at the footpoint and apex, respectively. The dashed and solid lines represent the trend of $\cos\theta$- and $(1-\theta/90)$-dependence, respectively.}
\label{fig:vmax}
\end{figure}

\figref{fig:spect_los1} presents the two extreme cases at the loop apex and footpoints\footnote{we truncated $2a/\sin30\deg=4a$ off the  loop ends, where rays with LOS angle of $30\deg$ would traverse only part of the loop cross-section and will contain an edge effect; therefore, we refer to the positions of $z=4a, L_0-4a$ as footpoints.}. At the footpoints, the intensity and line width variations do not change significantly with varying LOS angles, this is because the wavelength of $n=1$ standing slow mode is significantly longer than the loop radius. Thus, rays of any LOS angle are less likely to traverse both the node and anti-node. This is contrary to the short wavelength case for the fast sausage mode \citep{antolin2013}. In the short wavelength limit, LOS rays would traverse more fine structures along both radial and longitudinal directions, therefore the intensity and line width modulations are more complicated. However, in our case, the modulation of Doppler shift is significantly affected by LOS angles, since the longitudinal velocity dominates the perturbation. In a rough approximation, the amplitude of Doppler shift could be linearly de-projected by a factor of $1-\theta/90\deg$ or $1-2\theta/\pi$ , rather than $\cos\theta$ (\figref{fig:vmax}). A $\cos\theta$ de-projection could overestimate the velocity perturbation by a factor of more than 10\%. This empirical formula could be used in observations. The reason for the deviation from simple geometric projection is that each fluid element is projected by $\cos\theta$, while the overall Doppler-shifted spectrum including the contribution of all fluid elements along LOS does not necessarily follow the same trend. This result will affect the estimate of realistic longitudinal velocity perturbation with a two dimensional observation and the associated wave energy budget. The LOS effect may also lead to incorrect longitudinal mode identification \citep[see, e.g.,][]{antolin2011}.

\begin{figure*}[ht]
\centering
\includegraphics[width=0.6\textwidth]{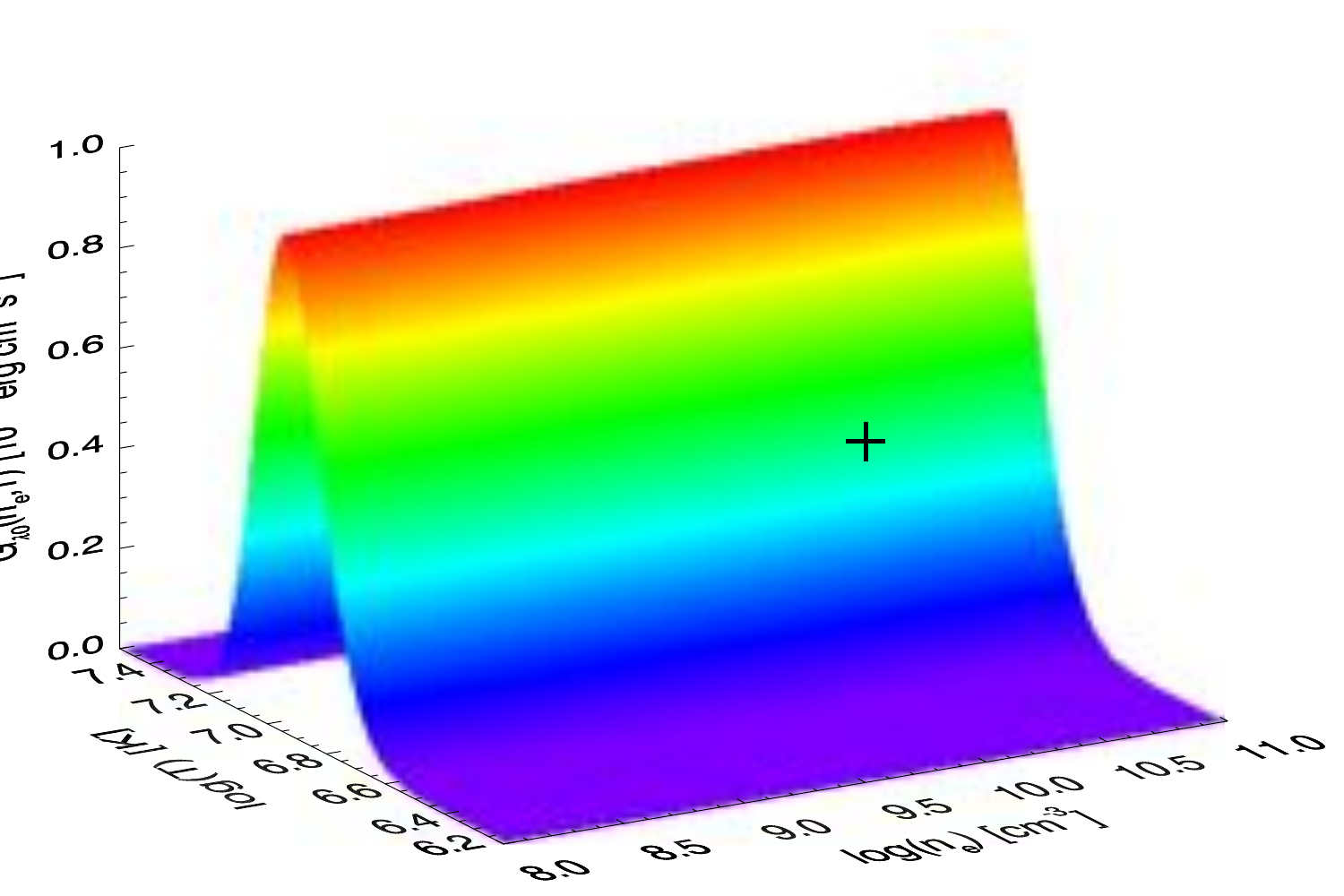}
\caption{The contribution function $G_{\lambda_0}(n_\e,T)$ for the \ion{Fe}{19} $\lambda1118.1\unit{\AA{}}$ line.  The maximum formation temperature is $\log T=6.95$ ($T=8.9\unit{MK}$). The plus symbol marks the region with the used loop parameters $[n_{\e\i},T_\i]$.}
\label{fig:goft_surf}
\end{figure*}

\textbf{The Doppler shift modulation exhibits similar patterns due to LOS variation at both the apex and footpoints (\figref{fig:spect_los1}{b, e}). Intriguingly, we found that positive modulation of the intensity is overwhelmingly in excess over negative modulations at every LOS angle (\figref{fig:spect_los1}{a} and \figref{fig:spect1_45} and \ref{fig:spect1_90}). This effect may halve the periodicity at regions where the intensity modulation is relatively weak, e.g., at the apex (\figref{fig:spect1_90}{d}).} This is not introduced by asymmetry in the geometry nor the distribution of electron density $n_e$ or plasma temperature $T$. \figref{fig:goft_surf} presents the contribution function of $\feline$, the plus sign indicates the loop parameters. $G_{\lambda_0}(n_e,T)$ varies less than 1\% with the electron density $n_e$ at high temperatures, but strongly peaks in temperature. In our case, positive temperature modulation leads to larger increase in $G_{\lambda_0}$ than the same amount of negative temperature modulation would do, therefore we have excess intensity enhancement when temperature increases. This was also found, although not mentioned, in the case of intensity variation in 193\AA{} bandpass \citep{antolin2013}. In contrast, this effect is missing in the 171\AA{} bandpass, as the contribution function $G_{171}$ has almost equal gradient with respect to $T$ in the temperature of interest, see Figure 6 in \citet{antolin2013}. This effect leads to a \textbf{halving} of periodicity in intensity and line width, especially in small LOS angles. It would become more significant for a loop with a temperature such that $\partial G_{\lambda_0}^2/\partial T^2$ reach extreme values. In such conditions, this effect would spread to broader spatial regions and larger LOS angles. We also notice that this effect would lead the asymmetry in emission intensity modulation (\figref{fig:spect_los1}{a}), however, it is highly likely to be neglected or removed by the data processing technique, e.g., detrending, running difference.  

\subsubsection{Longitudinal overtones}
We perform sit-and-stare mode observations for $n=2,3$ modes as well (Figures are not shown here). In spectroscopic observations, only a small segment of the spatial distribution of emission intensity and Doppler shift along a loop is normally measured. Without spatial information, it is not possible to judge the longitudinal mode number. However, with imaging observations, longitudinal overtones may be observed (see the next section).

\subsection{Imaging observation}
\label{sec:result_imager}
\begin{figure*}[ht]
\centering
\includegraphics[width=0.32\textwidth]{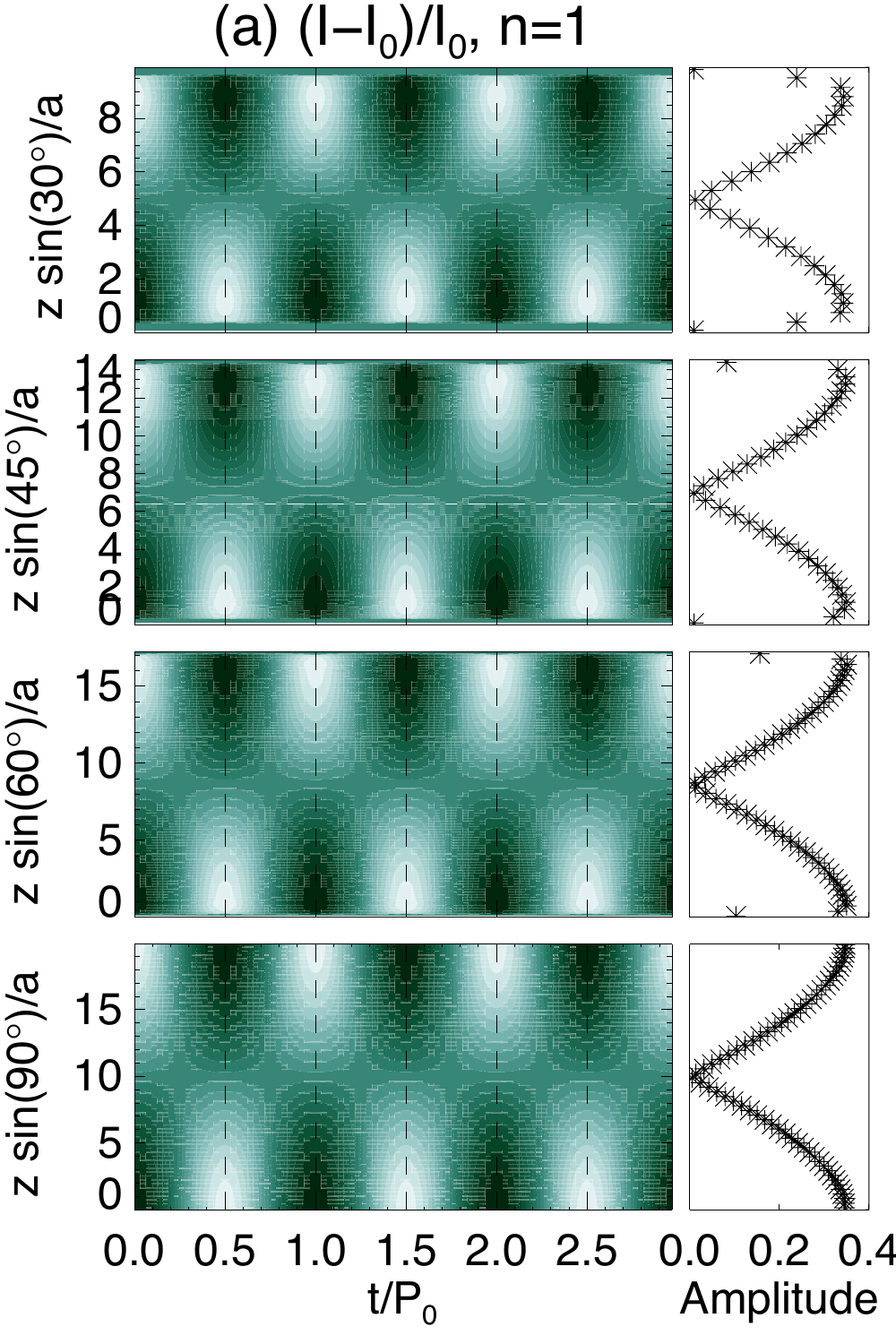}
\includegraphics[width=0.32\textwidth]{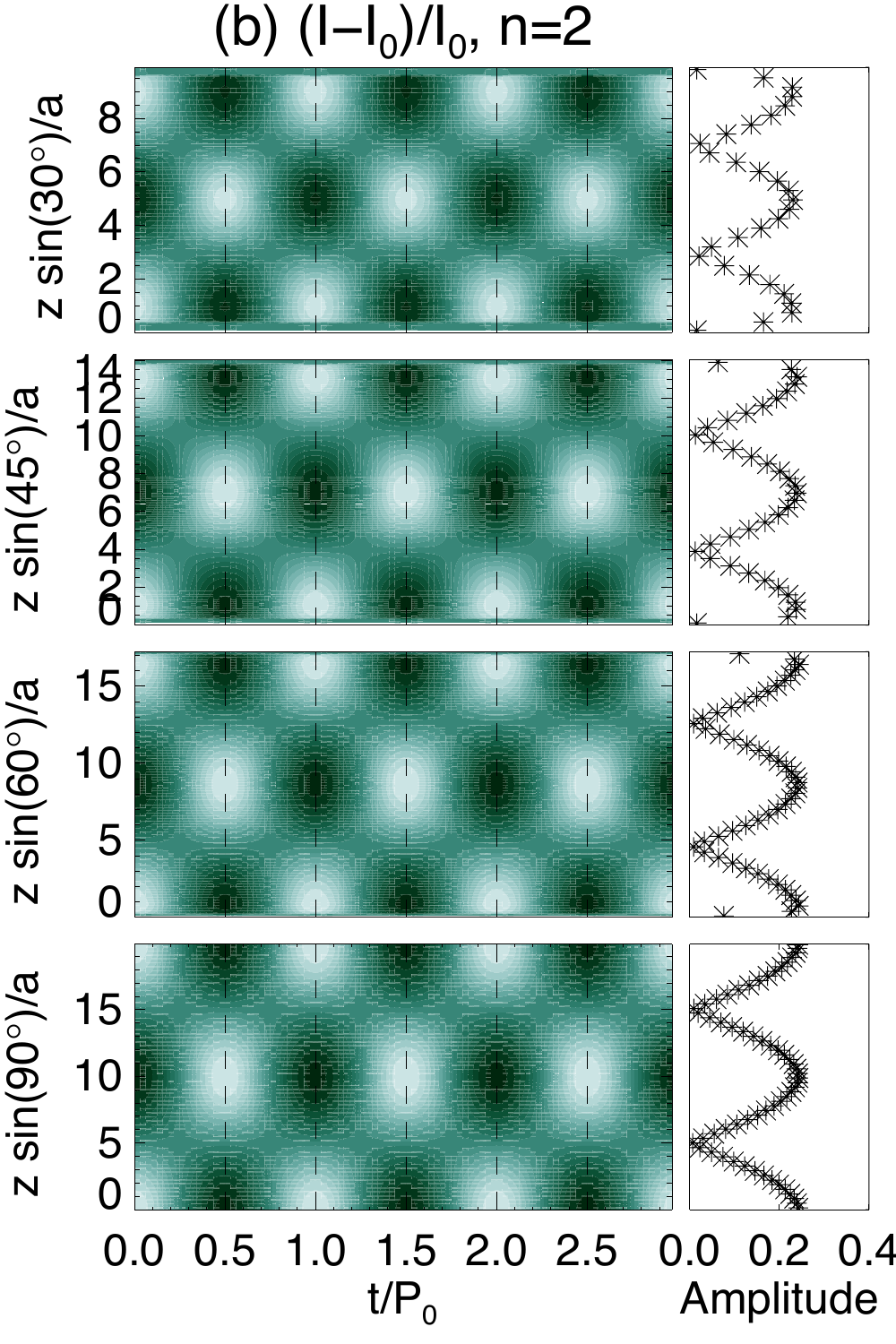}
\includegraphics[width=0.32\textwidth]{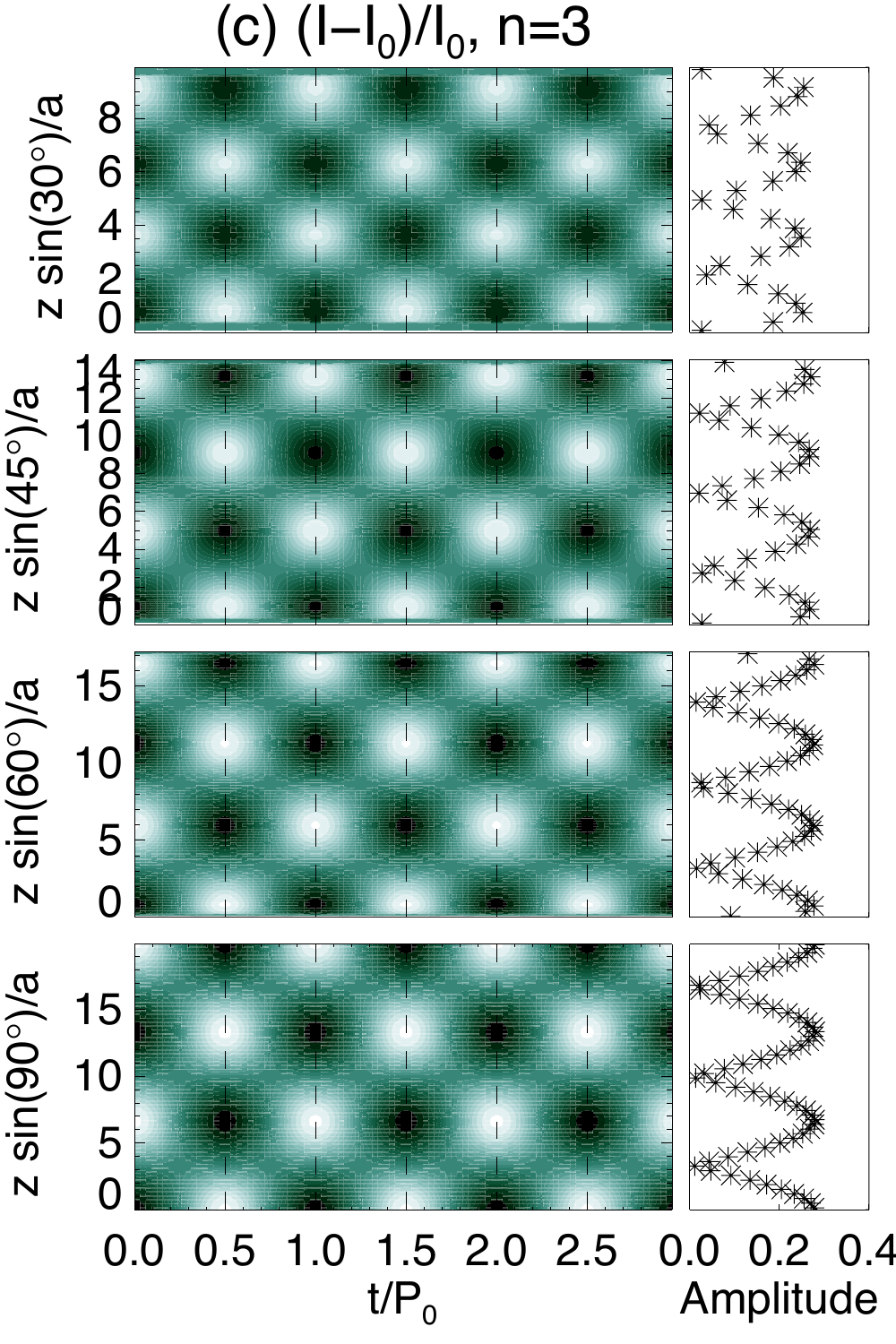}
\caption{Baseline ratio time-distance plot of AIA 94 \AA{} emission intensity along the central axis of the loop for modes $n=1$ (a), $2$ (b), $3$ (c), respectively.}
\label{fig:aia}
\end{figure*}

In imaging observations, the spatial distribution of the emission intensity is obtained. One could easily follow the spatio-temporal variation of a loop oscillation using the time-distance method \citep[see, e.g., ][]{yuan2012sm}. \figref{fig:aia}{a} shows the baseline-ratio difference plot of $n=1$ mode along the loop. The intensity perturbation is more significant at the loop footpoints than the loop apex. By changing the viewing angle, the intensity modulation changes slightly, but the overall spatio-temporal pattern remains unchanged. This is in agreement with the spectroscopic model, see discussions in \secref{sec:result_spect}. The time-distance plot along a loop could reveal the nodal structures of the longitudinal overtones (\figref{fig:aia}{b,c}. By comparing the amplitude variation along the loop, one could measure the longitudinal mode number $n$, and hence the wavelength of the slow mode $2L_0/n$. 

\subsection{Observing standing slow mode in hot loops with different temperatures}
\label{sec:temp}
\textbf{
So far, we only studied a $6.4\unit{MK}$ loop with fixed density and temperature ratio. As long as the background emission is insignificant compared to loop emission, the current result remains valid. However, if a coronal loop is heated to the hotter wing ($T_i>8.8\unit{MK}$) of $G_{\lambda_0}$ (\figref{fig:goft_surf}), the background emission could become stronger than the loop itself. In such cases, the observed Doppler shift and intensity modulation would be rather different. The free parameters are the loop temperature, density, the temperature and density ratio. It is not practical to iterate all possible combinations, so we only present possible scenarios as how to understand a slow standing mode properly. We vary the loop temperature, while keeping other free parameters unchanged. Several loop models are set up at the various temperatures where the SUMER \feline responses are signficant (see \tabref{tab:loop}).}

\begin{figure*}[ht]
\centering
\includegraphics[width=0.8\textwidth]{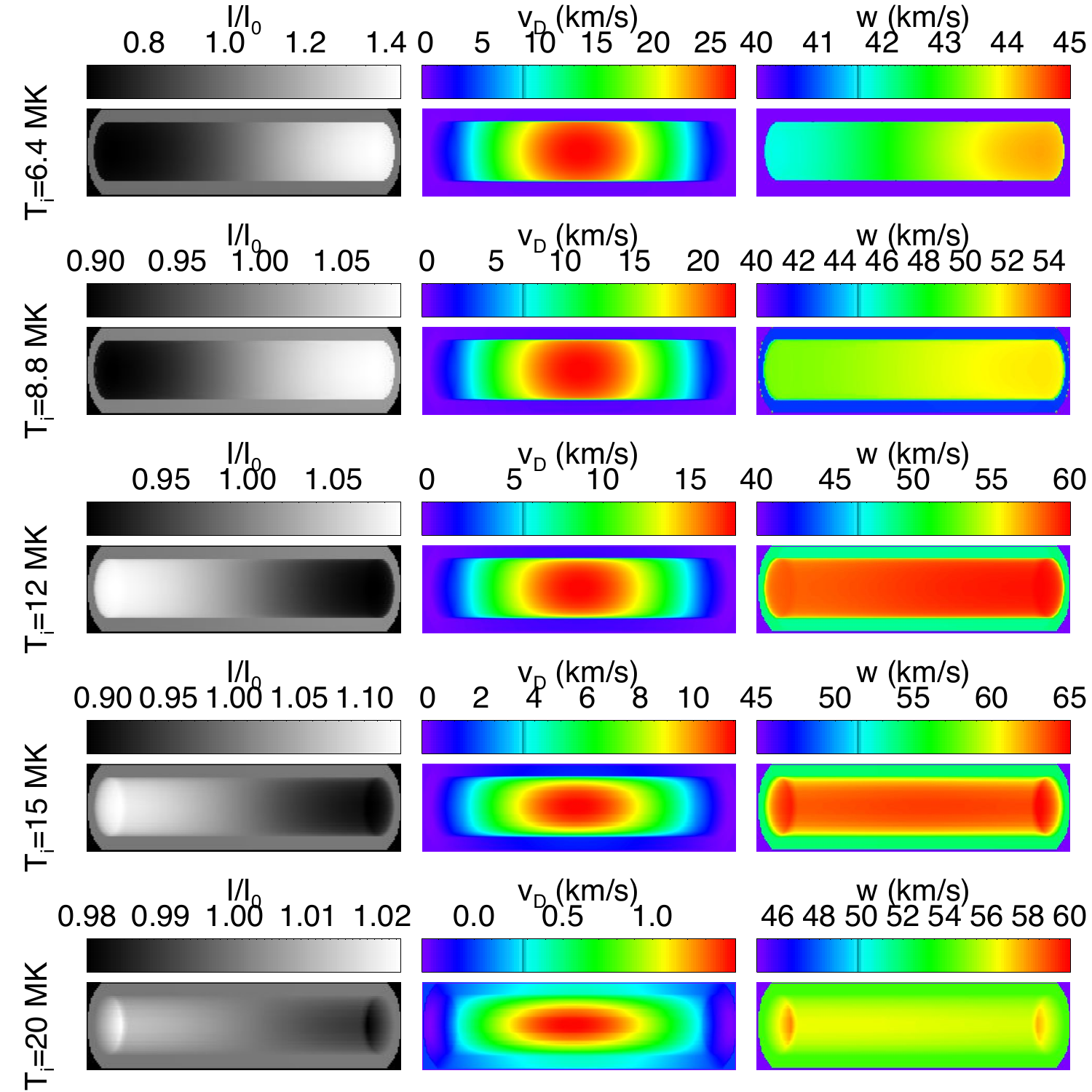}
\caption{Snapshots ($\theta=45\deg$, $t=P_0/8$) of the relative intensity emission (left column), the Doppler shift velocity (middle column), and the line width (right column) for loops at $T_i=6.4,8.8,12,15,20\unit{MK}$, respectively.}
\label{fig:syn}
\end{figure*}

\textbf{
\figref{fig:syn} illustrates snapshots of the relative intensity, Doppler velocity and line width of standing slow modes observed in loops at $T_i=$6.4, 8.8, 12, 15, and 20\unit{MK}, respectively. The snapshots are taken at $t=P_0/8$ at a viewing angle of $\theta=45\deg$ (Other viewing angles give similar results, therefore they are not shown). At the cooler wing of \feline, positive temperature modulation would enhance the intensity, while at the hotter wing, the opposite would occur. The intensity modulation becomes relatively small, as the ambient plasma emission intensity become significant or stronger than the loop itself (e.g, $T_i=20\unit{MK}$).
}

\begin{figure}[ht]
\centering
\includegraphics[width=0.5\textwidth]{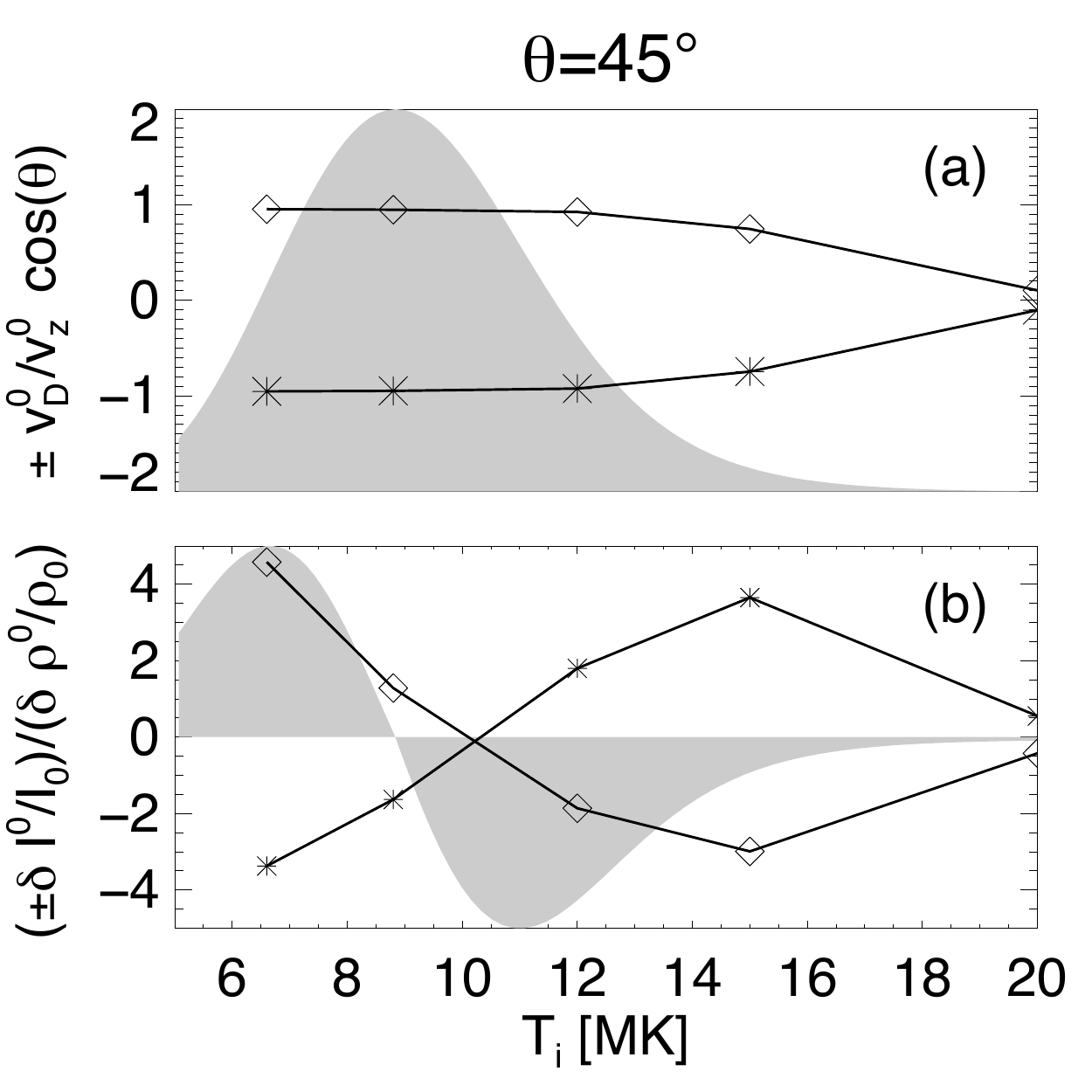}
\caption{For loops at different temperatures, (a) presents the normalized doppler velocity at a viewing angle of $
\theta=45\deg$ at loop apexes; (b) shows the normalized relative intensity modulation at loop footpoint, where $\delta I^0/I_0=(I_{t=0}-I_{t=P_0/4})/I_{t=P_0/4}$ and $-\delta I^0/I_0=(I_{t=P_0/2}-I_{t=P_0/4})/I_{t=P_0/4}$.  
 The shade areas illustrate the scaled $G_{\lambda_0}(T)$ (a) and $\partial G_{\lambda_0}(T) /\partial T$ (b) at $n_e=8.5\cdot10^9\unit{cm^{-3}}$, respectively. The diamond and asterisk denote, respectively, the positive and negative modulations for both Doppler shift and relative intensity. \label{fig:Tevar}}
\end{figure}

\textbf{
\figref{fig:Tevar}{b} shows the normalized relative intensity modulation $(\delta I^0/I_0)/ (\delta \rho/\rho_0)$, which is usually assumed to be 2 \citep{wang2009b}, at the loop footpoint against loop temperature. It illustrates how the sign of $\partial G_{\lambda_0}(T) /\partial T$ would modulate the intensity variation and causes the asymmetry effect (see discussions in \secref{sec:los}). We found that $(\delta I^0/I_0)/ (\delta \rho/\rho_0)$ could be considerably larger than 2, and that it is only close to 2 at regions where $|\partial G_{\lambda_0}(T) /\partial T|$ is relatively small. At extremely hot loops (20\unit{MK}), the normalized intensity modulation could even approach zero.}

\textbf{
On the other hand, the Doppler shift velocity of slow standing mode remains detectable for most loop temperatures. However, the normalized amplitude $v^0_D/(v^0_z \cos\theta)$ at the loop apex (it is symmetric for both positive and negative motions) deviates more and more from unity for loops at higher temperatures (\figref{fig:Tevar}{a}). This does not mean that the wave energy is undetected. It is still buried in the line width (\figref{fig:syn}, right column) or the skewness of the spectra, which are not measured in this study.
}

\section{Conclusion}
\label{sec:disc}
In this paper, we performed forward modelling to predict the observational \textbf{features} of standing slow modes in hot coronal loops. By considering the geometrical and instrumental effects, we measure the plasma emission intensity, Doppler shift, line width and spectrum modification caused by the standing slow mode. 

We found that the Doppler shift velocity is significantly affected by LOS effect, it could become undetectable at an viewing angle of $\theta=90\deg$. We found that a linear de-projection by the LOS angle is more accurate than the $\cos\theta$ de-projection as traditionally used in \citet{wang2011}. The emission intensity perturbation is normally a quarter-period out of phase with the Doppler shift velocity variation, both in time and space. This effect has been used to identify the standing slow wave mode. Positive temperature variation introduces more emission intensity enhancement than the same amount of negative temperature variation would reduce it, because $\partial G_{\lambda_0}/\partial T |_{T_0^+}$ is significantly larger than $\partial G_{\lambda_0}/\partial T |_{T_0^-}$, i.e., $\partial^2 G_{\lambda_0}/\partial T^2 |_{T_0}$ reaches positive extremes. This effect could lead to the \textbf{halving} of the periodicity in intensity and line width at the loop apex. \textbf{Half} periodicity could be also reached if the $\partial G_{\lambda_0}/\partial T |_{T_0^-}\gg\partial G_{\lambda_0}/\partial T |_{T_0^+}$, which could be only found at the hotter wing of $G_{\lambda_0}$, meaning that the loop has to be heated to a few tens of MK in mega flares. This second-order effect in the contribution function could also lead to asymmetry in the emission intensity modulation.

Spectroscopic observations with a sit-and-stare mode alone are not able to resolve longitudinal overtones along the loop, due to the lack of spatial information. With EUV imagers, the longitudinal overtones could be resolved by investigating the spatial distribution of the emission intensity modulation. A good way of studying standing slow modes is to use joint observations of spectrographs and imagers. 

The Doppler shift oscillation of a standing slow wave strongly depends on the viewing angle, for those observations with LOS angle close to $90\deg$, we may not detect Doppler shifts. For loops on the solar disk, a sit-and-stare campaign of spectrographic observation is favoured to be placed slightly off the loop apex, because there is a higher probability that the viewing angle would be close to $90\deg$. For spectrographic observations off-limb, most loops apexes are well exposed for observation at good viewing angles. 

\textbf{The emission contribution function plays a signficant role in determining the observational features of a standing slow wave. It may cause asymmetric intensity modulation for positive and negative temperature perturbations. This effect could be the opposite for loops at the cooler and hotter wings of the \feline line. The normalized relative intensity modulation $(\delta I^0/I_0)/ (\delta \rho/\rho_0)$ would be 2 only for the loops close to the peak response temperature of the \feline line. The Doppler shift velocity could be significantly smaller than the plasma velocity, if the background plasma emission becomes more significant or stronger than the loop emission. For loops at the hotter wing of a spectral line, the intensity modulation could be small, and one will only observe the Doppler shift. It may lead to a false interpretation of the result, e.g.,  in the case of a propagating fast wave \citet{tomczyk2007,vandoorsselaere2008}.
}

Imaging observations are subject to more variability, e.g., loop curvatures, plasma stratification, ambiguity of the loop footpoints. The amplitude variation along a loop is sufficient to identify the longitudinal overtones (\figref{fig:aia}). Normally the $n=1$ mode could be well identified even if the footpoints are hard to find. However, the $n>1$ overtones have shorter wavelength and the nodes close to the footpoints are less reliably measured, therefore, it poses some challenges to identify a $n>1$ overtone.

From the observables, we could attempt to perform MHD seismology \citep{wang2007}. For the input parameters, we allocate 10\% uncertainties to period $P_0$, loop length $L_0$. While we keep the perturbations of density and temperatures as the uncertainties. So we have loop length $L_0=100\pm10\unit{Mm}$, density $\rho_0=(1.4\pm0.2)\cdot10^{-11}\unit{kg\,m^{-3}}$ and temperature $T_0=(6.3\pm0.5)\unit{MK}$, then we estimate the magnetic field strength $B_0=(40.6\pm6.2)\unit{G}$ inside the loop. If we compare the result to the real input $40\unit{G}$, the main uncertainty is in the estimation of the parameters, while the assumption that $2L_0/P_0\simeq C_\T$ only result in about 1.5\% of uncertainty. In the long wavelength limit, this assumption remains valid and causes small uncertainties in MHD seismology. However, one has to be cautious in using this relation, if the wavelength is much shorter \textbf{or the plasma $\beta$ is not small, the associated relative error could be estimated by $(1+\gamma\beta_i/2)^{0.5}-1$.}

\acknowledgements
The research was supported by an Odysseus grant of the FWO Vlaanderen, the IAP P7/08 CHARM (Belspo), the Topping-Up grant Cor-Seis and the GOA-2015-014 (KU~Leuven), and the Open Research Program KLSA201312 of Key Laboratory of Solar Activity of National Astronomical Observatories of China (D.Y.). CHIANTI is a collaborative project involving George Mason University, the University of Michigan (USA) and the University of Cambridge (UK).

\bibliographystyle{apj}
\bibliography{yuan2015st}

\end{document}